\documentclass[manuscript,screen]{acmart}

\AtBeginDocument{%
  }

\setcopyright{acmlicensed}
\copyrightyear{2018}
\acmYear{2018}
\acmDOI{XXXXXXX.XXXXXXX}

\acmConference[Conference acronym 'XX]{Make sure to enter the correct
  conference title from your rights confirmation email}{June 03--05,
  2018}{Woodstock, NY}

\acmISBN{978-1-4503-XXXX-X/2018/06}

\usepackage{xcolor}

\newcommand{\citeg}[1]{\cite[e.g.,][]{#1}}
\usepackage{tabularx}
\newenvironment{quoteitalicized}
    {\begin{quote}}
    {\end{quote}}
\newcommand{\quotes}[2]{\begin{quoteitalicized}\textit{#1} -{#2}\end{quoteitalicized}}

\begin{document}

\title{Beyond Detection: Governing GenAI in Academic Peer Review as a Sociotechnical Challenge}


\author{Tatiana Chakravorti}
\email{tfc5416@psu.edu}
\affiliation{%
  \institution{Pennsylvania State University}
  \city{University Park}
  \state{Pennsylvania}
  \country{USA}
}

\author{Pranav Narayanan Venkit}
\authornote{These authors contributed equally to this research.}
\email{pnarayananvenkit@salesforce.com}
\affiliation{%
  \institution{Salesforce AI Research}
  \city{San Francisco}
  \state{California}
  \country{USA}
}

\author{Sourojit Ghosh}
\authornotemark[1]
\email{ghosh100@uw.edu}
\affiliation{%
  \institution{University of Washington}
  \city{Seattle}
  \state{Washington}
  \country{USA}
}

\author{Sarah Rajtmajer}
\email{smr48@psu.edu}
\affiliation{%
  \institution{Pennsylvania State University}
  \city{University Park}
  \state{Pennsylvania}
  \country{USA}
}

\renewcommand{\shortauthors}{Chakravorti et al.}

\begin{abstract}
Generative AI (GenAI) tools are increasingly entering academic peer-review workflows, raising questions about fairness, accountability, and the legitimacy of evaluative judgment. While these systems promise efficiency gains amid growing reviewer overload, their use introduces new sociotechnical risks. This paper presents a convergent mixed-method study combining discourse analysis of 448 social media posts with interviews with 14 area chairs and program chairs from leading AI and HCI conferences to examine how GenAI is discussed and experienced in peer review. Across both datasets, we find broad agreement that GenAI may be acceptable for limited supportive tasks, such as improving clarity or structuring feedback, but that core evaluative judgments, assessing novelty, contribution, and acceptance, should remain human responsibilities. At the same time, participants highlight concerns about epistemic harm, over-standardization, unclear responsibility, and adversarial risks such as prompt injection. User interviews reveal how structural strain and institutional policy ambiguity shift interpretive and enforcement burdens onto individual scholars, disproportionately affecting junior authors and reviewers. By triangulating public governance discourse with lived review practices, this work reframes AI-mediated peer review as a sociotechnical governance challenge and offers recommendations for preserving accountability, trust, and meaningful human oversight. Overall, we argue that AI-assisted peer review is best governed not by blanket bans or detection alone, but by explicitly reserving evaluative judgment for humans while instituting enforceable, role-specific controls that preserve accountability. We conclude with role-specific recommendations that formalize the support–judgment boundary.
\end{abstract}



\maketitle

\section{Introduction}

As generative AI (GenAI) is increasingly entering academic peer-review workflows, there is active discourse and a growing body of research around the opportunities for effective usage to reduce workloads on strained reviewers and increase academic equity \citeg{biswas2023chatgpt, chen2025envisioning, doskaliuk2025artificial}, as well as concerns around the impact of GenAI usage on the fairness, accountability, and legitimacy of scholarly evaluation. As debates continue, GenAI usage in peer review is already prevalent; \citet{liang2024monitoring} estimates that between 6.5\% and 17\% of reviews submitted to prominent NLP conferences during the 2023–2024 cycles contained AI-generated text, even where conference policies explicitly prohibited such use. In this environment, we explore the following research questions: 


    \begin{itemize}
    \item\textbf{RQ1}: How do peer-review stakeholders understand and negotiate appropriate use of GenAI in peer review?
    \item\textbf{RQ2}: What risks and harms do stakeholders observe or anticipate from GenAI?
    \item\textbf{RQ3}: How do governance practices and institutional factors shape how GenAI is adopted, contested, and regulated in peer review?
    \end{itemize}

Our study centers the perspectives of area/program chairs (ACs/PCs) of AI/HCI venues triangulated with public social media discourse, connecting debates about governance and legitimacy with labor conditions and decision-making practices. Our contributions frame AI-mediated peer review as a sociotechnical governance challenge rather than a tool adoption problem, revealing how policy ambiguity functions as a governance strategy. We highlight responsibility/risk redistribution to individual scholars and ways in which GenAI is reshaping peer review as relational and care work. 

\section{Background}
\subsection{Peer Review as a Value-Laden Practice}

Peer reviewing has been a cornerstone of the academic process for centuries, dating back to the editorial practices for medical journals in the 17$^{th}$ and 18$^{th}$ centuries, where editors would almost independently assess the quality of submitted research for publication \cite{drozdz2024peer, palmer1908editorial, shiflett1988difficult}. The modern model of academic peer review emerged in the 19$^{th}$ century, as editors began to acknowledge the limits of their own abilities and sought out others with some specialized knowledge required to evaluate a given submission \cite{burnham1990evolution, kronick1990peer}. The practice of peer reviewing was adopted by scientific journals (including ACM journals such as the \textit{Journal for the ACM}) in the mid-20$^{th}$ century, initially with the editor-only model before transitioning into the external-reviewers approach to account for growth and varying specializations. 

Peer reviewing research is more than evaluating science to pass muster for some conference/journal. Like much of the scientific process, peer reviewing is not value-neutral; it reflects and reinforces (and sometimes contests) disciplinary values about what and whose knowledges are considered legitimate, rigorous, and worth disseminating, through judgments about quality, novelty, and impact entangled with social norms, institutional incentives, and ethical commitments. Peer reviewing is a performance of power \cite{berkenkotter1995power, lipworth2011shifting} in which uncompensated reviewers can gatekeep the publication of knowledge to the scientific community. 
Authors are afforded little recourse, as resubmissions are more likely to face rejection \cite{stelmakh2021prior}. This can be especially true for a conference with as broad a call for papers as FAccT, which can result in individuals being assigned reviews beyond their expertise \cite{laufer2022four}. While FAccT and similar conferences try to account for such subjectivity with reviewer guidelines requiring decisions to be strongly motivated and consensual decision-making among anonymous reviewers \cite{rastogi2024randomized}, it is still possible that a value alignment between reviewers unfairly affects authors. 

Although prevalent practice in journals and conferences is double-blind peer review,
there is also ample evidence of peer reviewers exhibiting subconscious bias based on reviewers' perceptions of authors' identities along the lines of gender \citeg{borsuk2009name, helmer2017gender, pontille2014blind, whittaker2008journal}, nationality and English proficiency \citeg{marsh2008improving, tregenza2002gender}, and prestige and institutional affiliation \citeg{bornmann2010content, lee2013bias}. Anonymizing authorial (and sponsor/institution) identities during peer review can sometimes create other issues, such as potentially ethically-conflicted research appearing benign \cite{young2022confronting} and reviewers consciously or subconsciously trying to identify authors \citeg{chung2015double, haffar2019peer, pontille2014blind}. 

\subsection{GenAI in Peer Review: Approaches, Risks, and Concerns}

Peer reviewing as a practice has been radically impacted by the advent and popularity of GenAI tools, as their ability to generate natural language reviews for papers or respond to specific questions required during peer review means that reviewers can easily offload their tasks to these systems. There exist GenAI tools specifically for peer review, e.g., ReviewWriter \cite{su2023reviewriter}, which guides users through the writing process with intermittent nudges and instructions, without allowing them to upload a document and receive an AI-generated review\footnote{For information about other LLM-assisted generative review agents, see \citet{sun2025large}.}. This (and similar approaches \citeg{drori2024human, neshaei2024enhancing, thakkar2025can}) attempts to strike a balance in AI-assisted peer review, whereby humans write their own reviews with augmentation and support from GenAI, including review outlining, suggested line completion, dropdown lists of suggestions, and more, addressing issues of reviewer fatigue while still maintaining academic objectivity and rigor \citeg{checco2021ai, chidambaram2025the, iop2025, gruda2025three, robertson2023gpt4, zhuang2025large}. Other GenAI tools produce entirely synthetic paper reviews, whose capabilities including writing broad reviews \citeg{chitale2025autorev, gao2025reviewagents, taechoyotin2025remor, wang2020reviewrobot, zeng2025reviewrl}, adapting review text to the specific requirements of a conference \citeg{du2024llms, idahl2025openreviewer, zhou2024may}, or predicting peer review feedback and suggesting improvements \citeg{bougie2024generative, liu2023reviewergpt, sukpanichnant2024peerarg, tan2024peer, weng2025cycleresearcher}. 

In the field, one perspective views GenAI as a pathway for more equitable access into academia \cite[e.g.,][]{austin2023ai, berdejo2023ai, ghosh2024chatgpt, kohnke2025artificial, nixon2024catalyzing, roshanaei2023harnessing}, where students and researchers from communities historically marginalized in academia can leverage AI systems to participate in research activities at similar levels as their more privileged peers. This can take several forms in peer review, such as accurately understanding papers' contents by parsing the often-complicated lexicon of academic writing \cite[e.g.,][]{al2024navigating, ghosh2024chatgpt} and proofreading reviews written outside writers' first or native languages \cite[e.g.,][]{alalaq2024ai, sebastian2024artificial}. On the other hand, using GenAI tools for peer review raises serious concerns about the erosion of academic integrity \cite{liang2024monitoring}, over and above the fact that AI-generated reviews might contain hallucinations and otherwise shallow evaluations of rich scholarly content \cite{conroy2023chatgpt, chawla2024chatgpt, gao2025reviewagents}. The overuse of GenAI threatens an epistemic convergence as the low ceiling of originality in AI-generated text \citeg{bender2021dangers, birhane2022values} promises the reinforcement of traditional methods and established disciplinary norms. AI-generated reviews also tend to overinflate praise without substantive critique \cite{donker2023dangers, zhou2024llm}, miss potential errors made by researchers \cite{goldberg2024usefulness, liu2023reviewergpt}, and take up different points of view from human reviewers \cite{du2024llms, saad2024exploring, suleiman2024assessing, yu2024your}, thus contributing to higher rates of acceptance for poorer quality papers \cite{russo2025ai, thelwall2024fields}.  

A majority of research around the impact of GenAI on peer reviewing does not consider the direct perspectives of the area/program chairs and reviewers  (with a few exceptions \citeg{ali2025nursing, ebadi2025exploring}). Our work bridges this gap, centering their perspectives and contextualizing them within social media discourse. Our social media study captures public and semi-professional framings of GenAI in peer review, while interviews provide in-depth accounts of how reviewers and organizers understand and navigate these issues within their roles. Our interview protocol was developed independently and was not informed by the social media analysis, and both components were analyzed individually before putting the findings in conversation, using a convergent parallel mixed methods design \cite{demir2018convergent, kerrigan2014framework}. 

\section{Social Media Study}
We collected social media posts discussing the use of GenAI in peer review from LinkedIn, Reddit, and Twitter/X to capture diverse audiences and communication styles: LinkedIn for professional and academic discourse; Reddit for discussion-oriented and experiential exchanges; and Twitter/X for rapid, public-facing commentary. Using platform-specific search and filtering tools, we collected posts and comments containing keywords including "AI + peer review", "problems with AI peer review", "ChatGPT + peer review", and "LLM + peer review". Data collection focused on English language content between September 2025 and August 2026 and excluded private, locked, or deleted posts. Retweets without additional commentary and duplicate content were removed. Our final dataset consisted of 448 comments, approximately evenly distributed across these three platforms: 149 posts each from LinkedIn and Twitter/X; and, 150 from Reddit. We did not collect or attempt to identify individual identities. Our corpus is not intended to be representative of all platform discourse as it is a search-retrieved, English-language convenience sample shaped by our keyword queries and platform ranking algorithms. To improve coverage, we used multiple synonymous search strings across platforms and deduplicated results. Our findings are intended to be interpreted as prevalent framings within the retrieved corpus, not population prevalence estimates. 

We conducted qualitative discourse analysis \cite{zajda2020discourse, mogashoa2014understanding} on social media data to identify discussion patterns around AI in peer review, through an iterative coding process combining theory-informed deductive categories prevalent in scholarship on AI governance and academic evaluation, and inductive refinement based on recurring patterns.

\section{Interview study}
We conducted 14 semi-structured interviews \cite{varanasi2022feeling, thakkar2022machine} with top AI and HCI conference ACs and PC chairs to understand their perceptions, experiences of the current peer review process, as well as how and where AI can be useful and where it should not be used in peer review. 

\noindent \textbf{Recruitment.} We sent emails and social media messages on LinkedIn and Twitter/X to reach diverse academic researchers from AI/HCI conferences, including PhD students, post-doctoral researchers, assistant professors, and industry researchers. We recruited 14 participants (P1-P14) (see Appendix Table \ref{tab:subject1} for demographic information). Recruitment messages contained information about the study, expected interview length, compensation, inclusion criteria, and a link to the screener survey. We also employed snowball sampling to directly recruit ACs and PC chairs. 

\noindent \textbf{Interview protocol.} 30-60 minute semi-structured interviews \cite{adams2015conducting} were conducted virtually in September-November 2025. All interviews were recorded and transcripts for analysis were generated via Zoom. Theinterview protocol was designed to elicit participants’ experiences with peer review and their perspectives on the role of GenAI. We began by asking participants about their academic background, experience conducting peer reviews, and the tools or processes they typically use when assessing manuscripts. We invited them to describe whether and how they had used GenAI systems to support reviewing tasks, and which aspects of evaluation they believed these tools might enhance. We then explored perceptions of the benefits, risks, and appropriate boundaries of GenAI in peer review, asking about potential advantages, drawbacks, and areas where AI involvement might feel inappropriate or unsafe. Finally, we connected these experiences to broader ethical and governance concerns, prompting participants to reflect on guidelines, policies, or institutional norms they believe should shape the responsible use of GenAI in scholarly review. 

Interviews were thematically analyzed \cite{blandford2016qualitative, ding2022uploaders} as authors read transcripts multiple times before individually open-coding them and then grouping codes across authors into preliminary themes that reflected emergent conceptual groupings relevant to the research questions. Themes were refined through regular discussions among authors until they were finalized, as authors went back to the transcripts to verify that each theme was supported by rich data excerpts and sufficiently represented variations across participants.

We present findings from each dataset independently before synthesizing them through triangulation. We begin with the social media discourse analysis.

\section{Findings from Social Media: Six Themes}

\textbf{1. Trust Shifting Toward Procedural Mechanisms:}
Across the dataset, discourse was dominated by concerns about governance, enforceability, and verification, rather than purely evaluative judgments about whether AI is “good” or “bad.” The most frequent theme was trust shifting toward procedural mechanisms, present in 27.5\% of posts (123/448). Posts coded in this theme emphasized the need for formalized processes to manage uncertainty introduced by AI use, including calls for disclosure rules, institutional policies, enforcement mechanisms, and editorial or program committee oversight. These discussions often framed procedural safeguards as necessary to preserve the credibility of peer review and to restore confidence in the provenance and reliability of reviews.

{\textbf{2. Ethical Risk and c Harm:}
The second most prevalent theme was ethical risk and unequal harm, appearing in 23.2\% of posts (104/448). Ethical concerns clustered around confidentiality and privacy, the risk of hallucinated or incorrect statements, and broader worries about unfairness and bias. A subset of posts also raised concerns about uneven consequences and vulnerability, including the possibility that ambiguous rules or enforcement practices could disproportionately burden less powerful actors. Overall, ethical risk discourse was not limited to abstract principles; it frequently focused on concrete practices.

{\textbf{3. Boundary Setting around AI:}
A third major theme involved boundary setting, present in 21.7\% of posts (97/448). These posts drew explicit lines between acceptable and unacceptable uses of AI within peer review. Common acceptable uses included support tasks such as summarizing papers, improving grammar or clarity, structuring feedback, or helping reviewers organize their thoughts. In contrast, many posts rejected AI use for evaluative judgments, especially decisions about novelty, contribution, correctness, and accept/reject recommendations. This boundary was often framed as a defense of expert judgment and accountability, emphasizing that peer review requires contextual understanding and disciplinary responsibility that AI systems cannot reliably provide.

{\textbf{4. Responsibility and Care Work:}
The fourth most frequent theme was responsibility and care work, appearing in 17.2\% of posts (77/448). Posts coded in this theme portrayed reviewing as a professional obligation rather than a mechanical task. They highlighted accountability, diligence, and the value of constructive feedback. This discourse also included discomfort with the idea that AI could normalize minimal-effort reviewing or reduce peer review to a procedural formality. While responsibility language sometimes overlapped with policy debates, it was distinct in its focus on norms of scholarly conduct and the ethical meaning of reviewing labor.

{\textbf{5. Detection Culture and Policing:}}
Detection culture and policing of AI use appeared in 12.3\% of posts (55/448). These posts centered on claims that AI-written or AI-assisted reviews are recognizable and debated whether detection is feasible, fair, or desirable. Some argued for strict sanctions, bans, or the use of detection tools, while others warned about false positives and reputational harms. Across these discussions, a shared concern was that uncertainty about whether reviews were authored by humans undermines confidence in the process. Detection discourse, therefore, functioned less as a purely technical conversation and more as an indicator of shifting norms: provenance becomes a matter of suspicion and verification rather than assumed integrity.

{\textbf{6. Adversarial Peer Review and Prompt Injection}}
Discourse related to adversarial peer review and prompt injection were present in 10.3\% of posts (46/448). Posts in this category discussed tactics in which hidden or embedded instructions are inserted into manuscripts to manipulate LLM-based readers or reviewers (e.g., “prompt injection” using invisible text). This discourse framed peer review as entering a more adversarial ecosystem, where actors anticipate and exploit vulnerabilities introduced by AI-mediated reading and summarization. Discussions often extended from describing attacks to proposing countermeasures, such as stripping hidden text, filtering inputs, instituting procedural checks, or discouraging or prohibiting certain AI use cases.

\section{Findings from Interviews: {Six Dominant Themes}}

\subsection{Pre- and Post-AI Structural Strain in Peer Review}
All participants consistently mentioned peer review as a system that was already under significant strain, long before the emergence of GenAI. Across roles, they pointed to reviewer fatigue and overload, noting that a small pool of reviewers is repeatedly asked to evaluate a growing number of submissions. Reviewing was described as largely unpaid and undervalued labor, often added on top of heavy workloads that include teaching, research, service, and administrative responsibilities. Thus, reviewers feel stretched thin and have limited time and energy for each manuscript.
\quotes{It's challenging to find reviewers. Reviewers are really overloaded. All committee members have, like, 8 to 15 reviews that they're assigned to for one conference. This is a real problem of overload, which, I think, tempts people towards trying to use GenAI to reduce and manage their overloaded work.}{-P1}

P1 also added that these problems increase when you work as an AC or PC because, at that time, you need to check that your subcommittee members are not using AI for review. 
\quotes{GenAI use becomes especially risky at higher levels of conference leadership because when area chairs or subcommittee members use GenAI to draft meta-reviews, it can undermine the integrity of decisions that are supposed to reflect expert judgment}{-P1}

Participants also emphasized strong inconsistencies in review quality. While some reviews were described as careful, detailed, and constructive, others were brief and vague. This variability was linked not only to individual differences among reviewers but also to the lack of clear or shared reviewing standards. 
\quotes{Always, human reviews are not of high quality. Sometimes the reviews are short and low in quality.}{-P6}
Within this already fragile system, participants did not view GenAI as the root cause of these problems. Instead, they described GenAI as amplifying existing weaknesses. When reviewers are overloaded, pressed for time, and uncertain about standards, AI tools may appear attractive as a way to cope with demands. At the same time, participants worried that this could further normalize minimal effort reviewing and deepen existing quality gaps. P5 describes the huge number of submissions as a major problem these days, as people use GenAI to paraphrase and do multiple submissions.
\quotes{There has been a real uptake in the number of submitted papers. For example, with AAAI, I heard this year, 20,000 papers were submitted, and CHI were 6000. It's getting harder to tell what ethical submissions are. People were using AI to paraphrase their papers and submitting multiple copies of it.}{-P5}

P4 also added that as the number of submissions is getting higher day by day, it is difficult to find experienced reviewers. Getting reviewed by early-career PhD or master's students can produce more low-quality reviews. He also added details about how the use of generative AI is introducing challenges for senior reviewers. 

\quotes{At a senior role, I am on the opposite side when the other reviewers are saying that this is an AI-generated review or something. And also, we got directives that, if we are looking at other reviews as a senior reviewer. We need to make sure that this is not AI-generated. So, it now adds one more thing, we had to make sure that the reviews are high quality, the reviews are polite, the reviews are collegial, now we also have to make sure that the reviews are not AI-generated, because AI-generated reviews are also very polite. Like, most of the cases, that's what we have seen. So that makes it very hard to, hard to counter.}{-P4}

\subsection{Boundary Setting around AI}
Across interviews, 13 out of 14 participants, except P7, consistently engaged in what can be described as boundary setting around AI, actively defining which aspects of peer review could appropriately involve GenAI and which must remain the responsibility of human reviewers. Rather than expressing rejection of AI, most participants articulated clear limits on its use. AI was generally seen as acceptable for supportive or preparatory tasks, such as writing manuscripts, checking false references, improving grammar or clarity, organizing thoughts, or helping structure a review. These uses were framed as ways to reduce cognitive load without replacing the reviewer’s own reasoning.

\quotes{A lot of my students who are English-as-a-second-language speakers, their writing and their ability to express their ideas in English have gotten significantly better, and I know they're co-writing with AI because they tell me. They used to pay professional copyeditors to rewrite papers, now they don't need it.}{-P10}

P10 also mentions that GenAI can be super useful for checking grammar and typos, which can eventually save time for the reviewers, and they can use that time to check the technicality and ideas, as did P8. 
\quotes{I also have been using AI to proofread my language, and to correct grammar mistakes, and sometimes my review is too rambling, so I asked AI to shorten my writing without cutting any original meanings and writing that I had there, so that was very helpful for me to make sure my language is polished.}{-P8}

In contrast, participants drew firm boundaries around evaluative tasks. Uses of GenAI to assess a paper’s novelty, scientific contribution, methodological soundness, or overall merit, and especially to accept or reject recommendations, were widely described as inappropriate. Participants emphasized that these tasks require deep disciplinary expertise, familiarity with ongoing debates in the field, and sensitivity to context that cannot be reduced to patterns in training data. Many expressed concern that GenAI cannot understand why a contribution matters within scholarly conversations.

\quotes{LLMs or generative AI have been used extensively nowadays, but they should not replace human judgment. The final decision should be made by humans. LLMs can not accept or reject, or should not make any decisions in the future as well. When it goes to technical details they can not do it properly.}{-P12}

Importantly, these boundaries were not justified solely in terms of GenAI’s technical limitations. Instead, participants framed them in normative and professional terms. Judgment in peer review was described as inseparable from accountability; reviewers are expected to stand behind their evaluations, explain their reasoning, and take responsibility for the consequences of their recommendations. Delegating judgment to AI was seen as undermining this responsibility. 

\subsection{Ethical Risk and Unequal Harm}
Participants consistently raised ethical concerns about the use of GenAI in peer review, most commonly pointing to risks such as hallucinated or incorrect claims or false references, amplification of existing biases, misuse of AI, and potential breaches of confidentiality. All participants except P11 mentioned their concerns about the growing use of AI in Peer Review. Many worried that AI tools could introduce errors or misleading statements into reviews, particularly when reviewers rely on them without carefully verifying outputs. 

\quotes{These are unpublished works that are being fed into systems that can potentially ingest, maintain, and redistribute that information to other users or train themselves on it, so you end up kind of feeding in potentially invalid information. There are copyright issues involved in that as well. I just think it's really problematic. A review that I got on a paper that I wrote about, it was about family studies, ... And, one of the reviews mentioned that the paper was about children with ADHD. But it wasn't. There was no mention of ADHD or neurodivergence throughout the entire study, and this was a review that was recommending rejection. That made me wonder if maybe this was a detail that was hallucinated by GenAI?}{-P1}

P4 and P5 mentioned a critical ethical problem of duplicate submissions. They both came across duplicate submissions, as using GenAI, it is very easy to paraphrase. Therefore, authors are changing their paper writing and submitting duplicate papers. P4 also mentioned that AI can help to detect these duplicate submissions.

\quotes{One thing that AI can be useful for is selecting and flagging duplicate submissions. I have played senior roles in multiple conferences. I have noticed that there are duplicate submissions, and the duplicates are done in a very clever way using AI, so that it is not identical; the title will be slightly different, the abstract will be slightly different, and the introduction will be slightly different. It's almost deliberately meant to game the system, so everything will be slightly different, but they are basically the same paper.}{-P4}

Some participants expressed concern that AI systems may reproduce or intensify biases present in training data, shaping evaluations in ways that disadvantage certain topics, methods, or groups of scholars. But many participants are not very concerned about biases, as human review systems also have their own biases. Misuse of AI and its hallucination is a far more important concern mentioned by the participants. 
\quotes{An important concern is whether these AI-generated reviews are subsequently evaluated or verified by human reviewers. This is particularly critical due to the well-known issue of hallucinations in large language models. Without human oversight, it is unclear whether all the comments produced by the AI are valid, meaningful, or factually correct. Based on our own investigation of the AI-generated review for our paper, we found that some comments were not meaningful, and others were simply incorrect. This raises serious concerns about the reliability of AI-based reviews and highlights the need for human validation to ensure the accuracy and usefulness of the feedback.}{-P6}
P7 mentioned that this is gradually taking the critical thinking capabilities from humans, as peer review needs a lot of critical thinking. Beyond these general ethical risks, participants emphasized that the consequences of AI use would not be evenly shared across the academic community. Instead, they described a system in which some groups, particularly junior scholars, would bear a disproportionate share of the potential harm. Early-career authors were seen as especially vulnerable to low-quality or AI-assisted reviews, as they often have less power to contest unfair evaluations or request reconsideration. Participants noted that when feedback is vague, inaccurate, or inconsistent, junior authors may be more likely to internalize negative judgments or face career consequences without recourse. 

\subsection{Responsibility and Care Work}
This is also another emerging theme that came across the interviews, as 11 out of 14 participants mentioned this, except P5, P10, and P14. Participants frequently described peer review as more than a technical process for deciding which papers should be accepted or rejected. Instead, many framed it as a form of scholarly responsibility that involves mentorship, care for authors, and the reproduction of good work and professional norms within a field. Providing thoughtful feedback, explaining reasoning, and helping authors improve their work were described as central to what it means to be a good reviewer, particularly for early-career scholars.

\quotes{People who take it seriously and in good faith aren’t going to use [GenAI]… It’s really about finding people who are qualified to take it seriously. But I am afraid that people are gradually losing their responsibility and care for peer review.}{-P9}

Within this understanding, participants expressed concern that GenAI could gradually undermine the sense of responsibility associated with reviewing. Several worried that reliance on AI tools might normalize perfunctory or surface-level reviews, especially in an already overloaded system. When reviews are produced quickly or partially delegated to AI, participants feared that reviewers may feel less accountable for the content of their feedback, even if they technically approve or edit the final text.

\quotes{Previously, long reviews meant that the reviewer had thoroughly reviewed and done a great job, but now producing a longer review has become very easy… a long review doesn’t necessarily mean the reviewer put in a lot of work anymore.}{-P4}


\subsection{Strategic Ambiguity as Institutional Governance}
Across interviews, 9 out of 14 participants, P2, P3, P5, P8, P9, P10, P11, P12, and P14, pointed to the lack of clear, enforceable policies governing the use of GenAI in peer review. Participants explained that this ambiguity was motivated by several concerns. First, they noted that GenAI technologies are evolving rapidly, making it difficult to define stable rules that would remain relevant over time. Clear or rigid policies were seen as likely to become outdated quickly or to unintentionally prohibit legitimate uses. 
Second, participants described fears of institutional liability or controversy. Committing to firm positions was viewed as risky, especially in the absence of consensus across the scholarly community. By keeping policies vague, journals and conferences could avoid backlash from reviewers, authors, or professional societies. At the same time, participants recognized that this strategic ambiguity came with costs. When expectations are not clearly defined, responsibility for interpretation shifts downward to individual scholars. 
\quotes{There are copyright issues, there are ethical issues, and there’s a real concern about liability. If an institution makes a strong call and it turns out to be the wrong one, then someone has to take responsibility for that. And I think that’s why nobody really wants to be the one making that definitive decision.}{-P1}

Several participants noted that ambiguity also complicates enforcement. Without clear standards, it becomes difficult to determine when AI use crosses a line or warrants intervention. 
\quotes{If there isn’t a clear standard, then it becomes very hard to say when something actually crosses the line. You might feel that a review is AI-generated or inappropriate, but without a clear policy, it’s difficult to justify intervening or taking action, because there’s no agreed-upon boundary.}{-P14}

\subsection{Cross-Community Differences}

\subsubsection{HCI vs AI Reviewers}
Across interviews (P1–P14), we did not observe a sharp disciplinary split in participants’ overall stance toward GenAI in peer review. Participants across both HCI and AI communities broadly converged on core positions; peer review is already under strain, GenAI may be acceptable for limited supportive tasks, and evaluative judgment, e.g., assessing novelty, contribution, or accepting or rejecting recommendations, should remain a human responsibility. However, we did observe consistent differences in emphasis; that is, participants from HCI contexts tended to frame the problem in more relational and normative terms, whereas participants from AI contexts more often framed it as an infrastructural and governance challenge shaped by scale.

HCI participants more frequently described peer review as a social and moral practice, emphasizing mentorship, constructive feedback, and the relational responsibilities of reviewing. 
In contrast, AI participants more often situated the issue within the realities of high-volume conference reviewing and severe reviewer scarcity. Their accounts emphasized the scale and throughput pressures that make AI assistance important, alongside pragmatic questions about feasibility; what policies can realistically be enforced, how to manage disclosure, whether detection is workable, and how procedural systems might be designed to protect integrity without collapsing under administrative burden.
\quotes{There’s a sense that reviewing is part of how we take care of each other in the community… it’s not just about saying yes or no.}{-P5, HCI researcher}

\quotes{This year they received almost 40,000 papers… and it’s hard to even get three human reviews for each paper. If you have two human reviewers, then the third one can be an AI reviewer… not for ratings, but at least to provide comments.}{-P6, AI researcher}

\subsubsection{Senior vs. Junior Reviewers}
Across the interviews, we observe a clear difference in how senior reviewers and more junior or early-career reviewers describe what is at risk in AI-mediated peer review. Senior voices tend to frame the problem as one of system governance and meta-review integrity; their attention is on whether the review pipeline can still reliably filter, weight, and synthesize evaluations under scale and AI assistance. 
P10 illustrates the “system reliability” framing through a concrete failure mode in AI-generated meta-review; when LLMs summarize discussions, they struggle to downweight the low-quality or unserious reviewer input. P10 describes frustration that there was no way to tell the system that this review is terrible, and it should not actually consider this review in the summary, because LLMs treat content in the context window as having roughly equal importance.
\quotes{One of the things the AAAI conference did do was they used the AIs to help template or draft meta-reviews. That was a summary of the discussions. I think it really failed there. I think it was really bad there, because…there was no way for me to go in and say, this review is terrible. Like, you should not actually consider this review in the summary.}{-P10, associate prof}

In contrast, early-career participants focused strongly on the labor demands of peer review and reviewing no longer being a space for learning. P5, a postdoctoral researcher, described a sharp rise in her reviewing workload, being asked to review an unusually large number of papers within a very short period of time. She emphasized that this increase could not be explained simply by her growing seniority, but reflected broader structural pressures on the review system. P5 also highlighted how increased submission volume has weakened existing mentorship and quality-control practices. 
\quotes{There’s such a shortage of reviewers, we’re reaching out to junior people, who may not have the experience or guidance to know what’s acceptable. First-year PhD students reviewed before, but usually under guidance… now, because of sheer numbers, it’s impossible to do that kind of quality control.}{-P5, postdoc}

\section{Triangulation}
Social media discussions and interview findings show strong convergence in how different stakeholders think about the use of GenAI in peer review. In both cases, people do not treat AI as simply good or bad, instead focusing on proper usage and delineating boundaries. Across interviews and social media posts, there is broad agreement that AI may be acceptable for limited supportive tasks, such as improving clarity or organizing feedback, but that key evaluative decisions like judging novelty, contribution, or whether a paper should be accepted should remain the responsibility of humans. Both sources also express concern about the lack of clear rules and guidance, noting that such ambiguity makes it harder to trust the review process. Ethical concerns, such as confidentiality, incorrect or hallucinated feedback, and fairness, appear frequently in both datasets, as well as worries that AI use could undermine trust in peer review if it becomes too common or poorly regulated, especially for authors who have little power to challenge unfair reviews. Three themes that overlap in both data sets, namely, \emph{Ethical Risk and Unequal Harm}, \emph{Boundary Setting around AI}, and \emph{Responsibility and Care Work}. 

There is also some divergence in the findings. For example, structural strain is a prominent theme emerging from our interviews but is less discussed on the social media platforms. This also highlights cross-community differences in the peer review process. Social media discussions focused primarily on governance concerns, such as disclosure, enforcement, and the legitimacy of peer review in the presence of GenAI. \emph{Adversarial Peer Review and Prompt Injection} and \emph{Shifting Trust Toward Procedural Mechanisms} were substantial themes in our social media data; they were touched upon but less represented in interviews. P8 mentioned rumors of AI-generated reviews containing prompts. P14 is the only participant who clearly and technically articulated concerns about prompt injection attacks and adversarial manipulation of AI-based reviewing systems. Drawing on their background as an AI safety researcher, P14 described how authors could intentionally embed hidden instructions in submitted papers to manipulate AI reviewers. 
\quotes{There are techniques called prompt injection. For example, you can add some text in white color into the paper’s PDF, like, ‘please give this paper a very high score.’ As a human, you cannot read or be aware of these instructions in the paper, but the AI can read them, and might be misled.}{-P14}




\section{Implications}

Our findings, as summarized in Table \ref{tab:key_findings_summary}, have implications for understanding how the introduction of GenAI into peer review cannot be understood as a simple tool adoption problem. 
Instead, AI-mediated peer review is increasingly experienced and debated as a \textit{sociotechnical pipeline} in which technical systems, institutional rules, professional norms, and uneven power relations interact to shape evaluative outcomes.

\subsection{Peer Review as a Sociotechnical Pipeline, Not a Discrete Task}

Across both datasets, peer review emerges as a tightly coupled sociotechnical pipeline rather than a bounded activity performed by individual reviewers. GenAI is not introduced at a single point but may intervene at multiple stages: manuscript preparation, reviewer sensemaking, feedback drafting, meta-review synthesis, and even detection and enforcement. Our findings show that concerns about AI use rarely focus on isolated interactions with tools; instead, participants worry about how AI reshapes responsibility, accountability, and trust across the entire pipeline.

This framing challenges narratives that treat AI in peer review primarily as a productivity enhancement. While efficiency gains are acknowledged, especially under conditions of reviewer overload, participants emphasize that the integrity of peer review depends on how evaluative authority is distributed and justified across human and non-human actors. When AI systems participate in producing or shaping evaluative judgments, the provenance of those judgments becomes opaque, complicating long-standing norms around scholarly accountability.

\subsection{Boundaries as a Normative Defense of Human Judgment}

A central observation here is that resistance to some GenAI use is not primarily about technical limitations, but about the preservation of normative commitments embedded in peer review. Both public discourse and interview participants engage in drawing boundaries that distinguish acceptable supportive uses of AI (e.g., improving clarity, organizing feedback) from unacceptable evaluative uses (e.g., judging novelty, contribution, or acceptance). Clearly defining these cases is required and necessary to create a safer and more practical application of AI in peer review. 

These boundaries reflect an understanding of peer review as a moral and professional practice rather than a purely informational one. Evaluative judgment is tied to expertise, accountability, and the ability to justify decisions to affected parties. Delegating these judgments to AI systems is perceived as undermining the ethical foundation of peer review, even when AI outputs appear plausible or polished. Importantly, these boundaries are socially negotiated rather than fixed, highlighting the need for governance mechanisms that make such distinctions collectively enforceable.

\subsection{Structural Strain as a Driver of AI Adoption}

Our interview findings foreground structural strain as a critical but under-recognized driver of AI use in peer review. Reviewer scarcity, increasing submission volumes, and declining mentorship capacity create conditions in which AI tools become attractive coping mechanisms. However, social media discourse rarely centers these labor conditions, instead focusing on governance, enforcement, and legitimacy.

This divergence has important implications. Without addressing structural strain, governance efforts that rely solely on restriction or detection risk misdiagnosing the problem. Participants do not describe AI as causing the erosion of review quality; rather, AI amplifies existing weaknesses in an already fragile system. This suggests that sustainable governance must attend to workload distribution, reviewer development, and institutional incentives.

\subsection{Uneven Risk and the Redistribution of Harm}

We highlight how AI in peer review creates uneven risks: junior authors are more vulnerable to vague or incorrect AI-shaped feedback due to limited power to contest decisions; junior reviewers face ambiguity about acceptable AI use while being disproportionately pressured to accept review requests; and senior reviewers and chairs inherit new forms of invisible labor, including monitoring AI usage in reviewing. These findings complicate universal claims about fairness or efficiency. AI does not simply introduce new risks; it redistributes existing ones along lines of seniority, institutional authority, and visibility. Any governance framework that fails to account for these asymmetries risks reinforcing, rather than mitigating, inequities in academic evaluation.

\begin{table*}[t]
\centering
\footnotesize
\setlength{\tabcolsep}{5.8pt}
\renewcommand{\arraystretch}{1.24}
\begin{tabularx}{\textwidth}{@{}p{0.32\textwidth}p{0.32\textwidth}p{0.32\textwidth}@{}}
\toprule
\textbf{Opportunities for GenAI in Peer Review} &
\textbf{Risks of GenAI in Peer Review} &
\textbf{Recommendations for GenAI Use} \\
\midrule

1. GenAI is viewed as useful for \textbf{supportive and preparatory tasks} such as improving clarity, grammar, tone, and structuring feedback, reducing reviewer cognitive load. 
&
\textbf{1.Epistemic Risk}: AI-generated reviews may be shallow, generic, or hallucinated, undermining meaningful evaluation and scholarly rigor.
&
1. Clearly distinguish \textbf{supportive uses} of AI from \textbf{evaluative judgment}, explicitly reserving decisions about novelty, contribution, and acceptance for humans. \\

2. GenAI is seen as a tool that can \textbf{lower participation barriers}, particularly for non-native English speakers and reviewers with limited time.
&
\textbf{2.Ethical and Accountability Risk}: Use of AI blurs responsibility for review content, making it unclear who is accountable for errors or harms.
&
2. Require \textbf{meaningful human oversight and responsibility}, where reviewers remain accountable for all feedback submitted under their name. \\

3. AI assistance is framed as potentially helping sustain peer review in a context of \textbf{reviewer overload and submission growth}.
&
\textbf{3.Care and Quality Risk}: Reliance on AI may normalize minimal-effort reviewing and erode peer review as a form of scholarly care and mentorship.
&
3. Reframe peer review as \textbf{care work}, emphasizing diligence, explanation, and constructive engagement rather than review length or polish. \\

4. Procedural mechanisms (e.g., disclosure rules, oversight) are seen as necessary to maintain trust as AI use becomes more common.
&
\textbf{4.Unequal Harm Risk}: Ambiguous rules and low-quality AI-assisted reviews disproportionately harm junior authors and early-career reviewers.
&
4. Develop governance frameworks that explicitly consider \textbf{uneven impacts} on junior scholars and provide recourse for contesting flawed reviews. \\

5. Public discourse frames AI governance as a matter of \textbf{procedural legitimacy} rather than technical capability alone.
&
\textbf{5.Adversarial Risk}: Prompt injection and manipulation of AI-based reviewers introduce new attack surfaces in peer review.
&
5. Treat AI-mediated peer review as a \textbf{potentially adversarial system}, incorporating safeguards such as input sanitization, disclosure norms, and incident-response procedures. \\

\bottomrule
\end{tabularx}
\caption{Synthesis of core findings across research questions, showing how opportunities and risks of GenAI in peer review motivate calls for clearer boundaries, accountability, and governance in AI-mediated evaluation.}
\label{tab:key_findings_summary}
\end{table*}

\section{Recommendations: Governing AI-Mediated Peer Review as a Sociotechnical System}

Our findings indicate that GenAI is being introduced into peer review under structural strain (e.g., reviewer overload, rising submissions) while institutions rely on strategic ambiguity to avoid committing to enforceable rules. This produces predictable failure modes: (1) reviewers use AI to cope with time pressure; (2) chairs inherit new monitoring labor without reliable standards (and face difficulty because AI-generated reviews can appear `polite' and plausible); (3) harms concentrate on those with least recourse (especially junior authors and junior reviewers); and (4) procedural trust becomes the dominant legitimacy mechanism, shifting peer review from an assumption of integrity to an ecology of verification. Our complete recommendations are present in Table~\ref{tab:actor_recommendations} in the Appendix, and we elaborate below.


\noindent \textbf{For Reviewers:} As primary producers of evaluative text, reviewers should \textit{treat GenAI as assistive scaffolding without delegating evaluation}, using GenAI for supportive tasks only (e.g., improving clarity, grammar, tone; outlining review structure; summarizing their own notes) and not for passing judgments of quality or novelty of the work. If at all using it for any evaluative work, reviewers should consider \textit{adopting verification norms for AI-assisted claims} about such content, e.g., if AI produces claims such as ``the method fails to control X" or ``the authors cite Y incorrectly," reviewers should manually verify the accuracy of such claims, without taking them at face value. Reviewers should also \textit{preserve confidentiality of manuscripts} and treat unpublished work as high-risk. They should avoid pasting or uploading unpublished manuscripts into third-party tools or popular chatbots, unless the venue explicitly approves that workflow (and the tool provides adequate contractual/privacy protections), and only look to institutionally-approved systems. 

Generally, reviewers should use GenAI to \textit{improve the communication of critique}, in light of our findings around participants valuing AI for politeness and brevity. This can be especially useful for tone calibration, ensuring reviews (especially negative ones) are measured. Above all, our findings reflect a concerted desire that \textit{reviewers make labor visible} in the age of AI-generated content, demonstrating to authors that they have spent the appropriate time and mental effort in preparing structured, evidence-linked feedback (e.g., major/minor concerns, and actionable revisions). 


\noindent \textbf{For Area/Program Chairs:} Area/Program Chairs should \textit{replace ``AI detection" with ``review quality adjudication" procedures} based on review quality rather than provenance speculation, triggering escalation (e.g., request revision, solicit an additional review, downweight) of reviews that are generic, ungrounded, or inconsistent with manuscript content, irrespective of human or AI-production. This \textit{``downweighting'' mechanism could also be institutionalized} across venues, with ACs/PCs having mechanisms to downweight unreliable/low-quality reviews in final decision-making. 

ACs/PCs should also \textit{provide reviewer-facing micro-guidance at the moment of review submission}, such as (i) a concise permitted/prohibited list of AI usage, (ii) a reminder about confidentiality, and (iii) a requirement that major critiques cite manuscript sections. This directly targets the shallow/hallucinated review risk while reducing interpretive burden on chairs. Further, this could be part of a larger effort to support junior reviewers, while also \textit{protecting them through ``safe harbor'' clarification channels} to address uncertainty about acceptable AI use among early-career reviewers. Venues could provide confidential guidance channels (e.g., ``ask the chair'' policy interpretations) for learning-oriented compliance, and emphasize that disclosure or clarification requests will not be penalized.

\noindent \textbf{For Conferences and Journals:} Our results suggest publication institutions should \textit{move from strategic ambiguity to ``adaptive specificity''} by defining a two-layered policy structure: a set of rigid principles (e.g., accountability, confidentiality, transparency, and proportional enforcement) and some operational guidance (e.g., permitted/prohibited uses, tool categories, disclosure requirements, and enforcement processes) that is adapted to each review cycle and emergent AI capabilities. Instead of banning AI usage completely in reviewing, policies should \textit{codify a boundary between support and judgment} by explicitly distinguishing between permitted support and restricted usage to generate decisions. 

Conference/journals should \textit{require minimal disclosures of AI usage to avoid unequal punishment}, given concerns about unevenly distributed harms of AI disclosure, focusing on improving governance instead of creating stigma by designing policies that support transparency and auditability such as bounded disclosure, where reviewers select from a short checklist of AI-assisted activities (e.g., ``grammar/tone,'' ``outline,'' ``none'') without requiring tool names or full prompts. Where provenance concerns arise, institutions should \textit{introduce due-process protections and proportional remedies} that prioritize corrective actions over punishment. This can be done by requesting review revisions, adding extra reviewers, or downweighting unreliable reviews. Severe sanctions should require clear evidence and procedural safeguards to reduce reputational harm from false accusations, consistent with detection-culture concerns in social media discourse. At the same time, \textit{adversarial risks should be treated as a part of review integrity, not edge cases}, and venues that allow AI tooling should incorporate basic integrity controls (e.g., PDF sanitization, detection of hidden text layers, and warnings about copying full manuscripts into LLMs). Even venues that prohibit AI use should acknowledge adversarial possibilities, because reviewers may still use tools informally. Finally, keeping in mind that overload is a driver of AI adoption and AI restrictions might simply relocate strain and incentivize covert use, policies should \textit{address structural strain as governance} through labor-facing reforms, including reviewer load caps, incentives/recognition for high-quality reviews, reviewer training programs, and expanded reviewer pools.

\noindent\textbf{For AI Tool Designers:} For designers of AI tools to be used in reviewing contexts, they should \textit{focus on accountable assistance that makes choices visible}, in a way that clearly delineates AI contributions through tracked suggestions, provenance markers, and editable drafts that require reviewer confirmation. Such tools must contain \textit{guardrails that operationalize the support/judgment boundary} through interaction patterns that discourage evaluative delegation (e.g., block generating accept/reject recommendations, require evidence citations for claims, and prompt users to add manuscript-grounded justification). They must also be robust enough to \textit{anticipate adversarial behavior and prompt injection}, detecting hidden layers, unusual formatting, or suspicious instruction-like text in PDFs, and surface appropriate warnings. Finally, given widespread confidentiality concerns, designers should \textit{build confidentiality-preserving workflows} and prioritize privacy-by-design: local/on-device processing when possible, contractual guarantees for non-retention/non-training, and clear notices about data handling. 

\section{Limitations}
This study has several limitations. First, all data are English language and largely reflect Western, primarily USA based academic contexts, which may limit applicability to other scholarly systems. Second, the social media dataset is not statistically representative; it is a keyword based, search retrieved convenience sample shaped by platform algorithms, and theme frequencies should not be interpreted as prevalence estimates. Third, findings are grounded mainly in AI and HCI conference contexts, and their transferability to journal review or disciplines beyond AI or HCI remains uncertain. Finally, data were collected during a specific period of rapid GenAI change, and norms and governance practices may evolve over time.

\section{Conclusion}
This study examined how GenAI is being used in academic peer review and how reviewers and chairs understand its role. Across interviews and social media discussions, we find broad agreement that GenAI may be useful for limited support tasks, but that key evaluative decisions such as judging novelty, contribution, and acceptance should remain human responsibilities. Participants also raised concerns about inaccurate feedback, unclear accountability, and new risks introduced by AI use. Our findings show that these concerns are shaped by existing pressures in peer review, including reviewer overload and unclear policies, which place greater burdens on individuals, especially early-career scholars. Together, these perspectives highlight that AI in peer review is not just a technical issue, but a governance challenge. We argue that responsible use of GenAI requires clear boundaries, meaningful human oversight, and attention to the labor and care involved in scholarly evaluation.

\newpage


\section*{Ethical Consideration Statement}
IRB approval for human subjects research was obtained prior to participant recruitment and data collection. Participants were fully informed about the nature of the study, potential risks, and their right to withdraw at any time without penalty. Interviews were conducted with sensitivity to participants’ professional roles, and no identifying information about specific conferences, incidents, or individuals was collected or reported. Data were anonymized, stored securely, and used only for the purposes described to participants.

Given that this work examines governance practices and potentially sensitive behaviors in academic peer review, we took care to minimize reputational and professional risk by focusing analysis on systemic patterns rather than individual compliance or misconduct. Public social media data were limited to publicly accessible posts, with no attempts made to identify or contact authors of the said posts. Our work directly addresses research integrity and transparency, with the aim of fostering more inclusive, accountable, and ethically grounded conversations about the use of generative AI in peer review.

\subsection*{GenAI Usage Statement}

The authors used ChatGPT only for minor grammar and sentence-structure checks. No generative AI tool was used to generate manuscript text, and the authors remain fully responsible for the originality of the submission.


\bibliographystyle{ACM-Reference-Format}
\bibliography{sample-base}


\begin{thebibliography}{93}


\ifx \showCODEN    \undefined \def \showCODEN     #1{\unskip}     \fi
\ifx \showISBNx    \undefined \def \showISBNx     #1{\unskip}     \fi
\ifx \showISBNxiii \undefined \def \showISBNxiii  #1{\unskip}     \fi
\ifx \showISSN     \undefined \def \showISSN      #1{\unskip}     \fi
\ifx \showLCCN     \undefined \def \showLCCN      #1{\unskip}     \fi
\ifx \shownote     \undefined \def \shownote      #1{#1}          \fi
\ifx \showarticletitle \undefined \def \showarticletitle #1{#1}   \fi
\ifx \showURL      \undefined \def \showURL       {\relax}        \fi
\providecommand\bibfield[2]{#2}
\providecommand\bibinfo[2]{#2}
\providecommand\natexlab[1]{#1}
\providecommand\showeprint[2][]{arXiv:#2}

\bibitem[Adams(2015)]%
        {adams2015conducting}
\bibfield{author}{\bibinfo{person}{William~C Adams}.} \bibinfo{year}{2015}\natexlab{}.
\newblock \showarticletitle{Conducting semi-structured interviews}.
\newblock \bibinfo{journal}{\emph{Handbook of practical program evaluation}} (\bibinfo{year}{2015}), \bibinfo{pages}{492--505}.
\newblock


\bibitem[Al~Sawi and Alaa(2024)]%
        {al2024navigating}
\bibfield{author}{\bibinfo{person}{Islam Al~Sawi} {and} \bibinfo{person}{Ahmed Alaa}.} \bibinfo{year}{2024}\natexlab{}.
\newblock \showarticletitle{Navigating the impact: a study of editors’ and proofreaders’ perceptions of AI tools in editing and proofreading}.
\newblock \bibinfo{journal}{\emph{Discover Artificial Intelligence}} \bibinfo{volume}{4}, \bibinfo{number}{1} (\bibinfo{year}{2024}), \bibinfo{pages}{23}.
\newblock


\bibitem[Alalaq(2024)]%
        {alalaq2024ai}
\bibfield{author}{\bibinfo{person}{Ahmed~Shaker Alalaq}.} \bibinfo{year}{2024}\natexlab{}.
\newblock \showarticletitle{AI tools assisting in the proofreading and scientific review of research papers}.
\newblock \bibinfo{journal}{\emph{ScienceOpen Preprints}} (\bibinfo{year}{2024}).
\newblock


\bibitem[Ali and Shaban(2025)]%
        {ali2025nursing}
\bibfield{author}{\bibinfo{person}{Sayed~Ibrahim Ali} {and} \bibinfo{person}{Mostafa Shaban}.} \bibinfo{year}{2025}\natexlab{}.
\newblock \showarticletitle{Nursing Academic Reviewers’ Perspectives on AI-Assisted Peer Review: Ethical Challenges and Acceptance}.
\newblock \bibinfo{journal}{\emph{International nursing review}} \bibinfo{volume}{72}, \bibinfo{number}{3} (\bibinfo{year}{2025}).
\newblock


\bibitem[Aly et~al\mbox{.}(2023)]%
        {aly2023changing}
\bibfield{author}{\bibinfo{person}{Mariam Aly}, \bibinfo{person}{Eliana Colunga}, \bibinfo{person}{MJ Crockett}, \bibinfo{person}{Matthew Goldrick}, \bibinfo{person}{Pablo Gomez}, \bibinfo{person}{Franki~YH Kung}, \bibinfo{person}{Paul~C McKee}, \bibinfo{person}{Miriam P{\'e}rez}, \bibinfo{person}{Sarah~M Stilwell}, {and} \bibinfo{person}{Amanda~B Diekman}.} \bibinfo{year}{2023}\natexlab{}.
\newblock \showarticletitle{Changing the culture of peer review for a more inclusive and equitable psychological science.}
\newblock \bibinfo{journal}{\emph{Journal of Experimental Psychology: General}} \bibinfo{volume}{152}, \bibinfo{number}{12} (\bibinfo{year}{2023}), \bibinfo{pages}{3546}.
\newblock


\bibitem[Austin et~al\mbox{.}(2023)]%
        {austin2023ai}
\bibfield{author}{\bibinfo{person}{Tasha Austin}, \bibinfo{person}{Bharat~S Rawal}, \bibinfo{person}{Alexandra Diehl}, {and} \bibinfo{person}{Jonathan Cosme}.} \bibinfo{year}{2023}\natexlab{}.
\newblock \showarticletitle{AI for equity: Unpacking potential human bias in decision making in higher education}.
\newblock \bibinfo{journal}{\emph{AI, Computer Science and Robotics Technology}} \bibinfo{number}{13} (\bibinfo{year}{2023}).
\newblock


\bibitem[Beecher and Wang(2025)]%
        {beecher2025peer}
\bibfield{author}{\bibinfo{person}{Kate Beecher} {and} \bibinfo{person}{Joshua Wang}.} \bibinfo{year}{2025}\natexlab{}.
\newblock \showarticletitle{Peer reviewer fatigue, or peer reviewer refusal?}
\newblock \bibinfo{journal}{\emph{Accountability in Research}} \bibinfo{volume}{32}, \bibinfo{number}{5} (\bibinfo{year}{2025}), \bibinfo{pages}{838--844}.
\newblock


\bibitem[Bel{\'e}m~de Oliveira(2024)]%
        {belem2024challenge}
\bibfield{author}{\bibinfo{person}{Jos{\'e} Bel{\'e}m~de Oliveira}.} \bibinfo{year}{2024}\natexlab{}.
\newblock \bibinfo{title}{The challenge of reviewers scarcity in academic journals: payment as a viable solution}.
\newblock \bibinfo{numpages}{eED1194}~pages.
\newblock


\bibitem[Bender et~al\mbox{.}(2021)]%
        {bender2021dangers}
\bibfield{author}{\bibinfo{person}{Emily~M Bender}, \bibinfo{person}{Timnit Gebru}, \bibinfo{person}{Angelina McMillan-Major}, {and} \bibinfo{person}{Shmargaret Shmitchell}.} \bibinfo{year}{2021}\natexlab{}.
\newblock \showarticletitle{On the dangers of stochastic parrots: Can language models be too big?}. In \bibinfo{booktitle}{\emph{Proceedings of the 2021 ACM conference on fairness, accountability, and transparency}}. \bibinfo{pages}{610--623}.
\newblock


\bibitem[Berdejo-Espinola and Amano(2023)]%
        {berdejo2023ai}
\bibfield{author}{\bibinfo{person}{Violeta Berdejo-Espinola} {and} \bibinfo{person}{Tatsuya Amano}.} \bibinfo{year}{2023}\natexlab{}.
\newblock \showarticletitle{AI tools can improve equity in science}.
\newblock \bibinfo{journal}{\emph{Science}} \bibinfo{volume}{379}, \bibinfo{number}{6636} (\bibinfo{year}{2023}), \bibinfo{pages}{991--991}.
\newblock


\bibitem[Berkenkotter(1995)]%
        {berkenkotter1995power}
\bibfield{author}{\bibinfo{person}{Carol Berkenkotter}.} \bibinfo{year}{1995}\natexlab{}.
\newblock \showarticletitle{The power and the perils of peer review}.
\newblock \bibinfo{journal}{\emph{Rhetoric review}} \bibinfo{volume}{13}, \bibinfo{number}{2} (\bibinfo{year}{1995}), \bibinfo{pages}{245--248}.
\newblock


\bibitem[Birhane et~al\mbox{.}(2022)]%
        {birhane2022values}
\bibfield{author}{\bibinfo{person}{Abeba Birhane}, \bibinfo{person}{Pratyusha Kalluri}, \bibinfo{person}{Dallas Card}, \bibinfo{person}{William Agnew}, \bibinfo{person}{Ravit Dotan}, {and} \bibinfo{person}{Michelle Bao}.} \bibinfo{year}{2022}\natexlab{}.
\newblock \showarticletitle{The values encoded in machine learning research}. In \bibinfo{booktitle}{\emph{Proceedings of the 2022 ACM conference on fairness, accountability, and transparency}}. \bibinfo{pages}{173--184}.
\newblock


\bibitem[Biswas et~al\mbox{.}(2023)]%
        {biswas2023chatgpt}
\bibfield{author}{\bibinfo{person}{Som Biswas}, \bibinfo{person}{Dushyant Dobaria}, {and} \bibinfo{person}{Harris~L Cohen}.} \bibinfo{year}{2023}\natexlab{}.
\newblock \showarticletitle{ChatGPT and the future of journal reviews: a feasibility study}.
\newblock \bibinfo{journal}{\emph{The Yale Journal of Biology and Medicine}} \bibinfo{volume}{96}, \bibinfo{number}{3} (\bibinfo{year}{2023}), \bibinfo{pages}{415}.
\newblock


\bibitem[Blandford et~al\mbox{.}(2016)]%
        {blandford2016qualitative}
\bibfield{author}{\bibinfo{person}{Ann Blandford}, \bibinfo{person}{Dominic Furniss}, {and} \bibinfo{person}{Stephann Makri}.} \bibinfo{year}{2016}\natexlab{}.
\newblock \bibinfo{booktitle}{\emph{Qualitative HCI research: Going behind the scenes}}.
\newblock \bibinfo{publisher}{Morgan \& Claypool Publishers}.
\newblock


\bibitem[Bornmann et~al\mbox{.}(2010)]%
        {bornmann2010content}
\bibfield{author}{\bibinfo{person}{Lutz Bornmann}, \bibinfo{person}{Christophe Weymuth}, {and} \bibinfo{person}{Hans-Dieter Daniel}.} \bibinfo{year}{2010}\natexlab{}.
\newblock \showarticletitle{A content analysis of referees’ comments: how do comments on manuscripts rejected by a high-impact journal and later published in either a low-or high-impact journal differ?}
\newblock \bibinfo{journal}{\emph{Scientometrics}} \bibinfo{volume}{83}, \bibinfo{number}{2} (\bibinfo{year}{2010}), \bibinfo{pages}{493--506}.
\newblock


\bibitem[Borsuk et~al\mbox{.}(2009)]%
        {borsuk2009name}
\bibfield{author}{\bibinfo{person}{Robyn~M Borsuk}, \bibinfo{person}{Lonnie~W Aarssen}, \bibinfo{person}{Amber~E Budden}, \bibinfo{person}{Julia Koricheva}, \bibinfo{person}{Roosa Leimu}, \bibinfo{person}{Tom Tregenza}, {and} \bibinfo{person}{Christopher~J Lortie}.} \bibinfo{year}{2009}\natexlab{}.
\newblock \showarticletitle{To name or not to name: The effect of changing author gender on peer review}.
\newblock \bibinfo{journal}{\emph{BioScience}} \bibinfo{volume}{59}, \bibinfo{number}{11} (\bibinfo{year}{2009}), \bibinfo{pages}{985--989}.
\newblock


\bibitem[Bougie and Watanabe(2024)]%
        {bougie2024generative}
\bibfield{author}{\bibinfo{person}{Nicolas Bougie} {and} \bibinfo{person}{Narimasa Watanabe}.} \bibinfo{year}{2024}\natexlab{}.
\newblock \showarticletitle{Generative adversarial reviews: When LLMs become the critic}.
\newblock \bibinfo{journal}{\emph{arXiv preprint arXiv:2412.10415}} (\bibinfo{year}{2024}).
\newblock


\bibitem[Brooks(2025)]%
        {brooks2025bad}
\bibfield{author}{\bibinfo{person}{Patrick Brooks}.} \bibinfo{year}{2025}\natexlab{}.
\newblock \showarticletitle{How bad gatekeepers undermine good science}.
\newblock \bibinfo{journal}{\emph{Synthese}} \bibinfo{volume}{206}, \bibinfo{number}{5} (\bibinfo{year}{2025}), \bibinfo{pages}{267}.
\newblock


\bibitem[Burnham(1990)]%
        {burnham1990evolution}
\bibfield{author}{\bibinfo{person}{John~C Burnham}.} \bibinfo{year}{1990}\natexlab{}.
\newblock \showarticletitle{The evolution of editorial peer review}.
\newblock \bibinfo{journal}{\emph{Journal of the American Medical Association}} \bibinfo{volume}{263}, \bibinfo{number}{10} (\bibinfo{year}{1990}), \bibinfo{pages}{1323--1329}.
\newblock


\bibitem[Chakravorti et~al\mbox{.}(2025)]%
        {chakravorti2025social}
\bibfield{author}{\bibinfo{person}{Tatiana Chakravorti}, \bibinfo{person}{Xinyu Wang}, \bibinfo{person}{Pranav~Narayanan Venkit}, \bibinfo{person}{Sai Koneru}, \bibinfo{person}{Kevin Munger}, {and} \bibinfo{person}{Sarah Rajtmajer}.} \bibinfo{year}{2025}\natexlab{}.
\newblock \showarticletitle{Social Scientists on the Role of AI in Research}. In \bibinfo{booktitle}{\emph{Proceedings of the AAAI/ACM Conference on AI, Ethics, and Society}}, Vol.~\bibinfo{volume}{8}. \bibinfo{pages}{528--540}.
\newblock


\bibitem[Chawla(2024)]%
        {chawla2024chatgpt}
\bibfield{author}{\bibinfo{person}{Dalmeet~Singh Chawla}.} \bibinfo{year}{2024}\natexlab{}.
\newblock \showarticletitle{Is Chat{GPT} corrupting peer review? Telltale words hint at {AI} use}.
\newblock \bibinfo{journal}{\emph{Nature}} \bibinfo{volume}{628}, \bibinfo{number}{8008} (\bibinfo{year}{2024}), \bibinfo{pages}{483--484}.
\newblock


\bibitem[Checco et~al\mbox{.}(2021)]%
        {checco2021ai}
\bibfield{author}{\bibinfo{person}{Alessandro Checco}, \bibinfo{person}{Lorenzo Bracciale}, \bibinfo{person}{Pierpaolo Loreti}, \bibinfo{person}{Stephen Pinfield}, {and} \bibinfo{person}{Giuseppe Bianchi}.} \bibinfo{year}{2021}\natexlab{}.
\newblock \showarticletitle{AI-assisted peer review}.
\newblock \bibinfo{journal}{\emph{Humanities and social sciences communications}} \bibinfo{volume}{8}, \bibinfo{number}{1} (\bibinfo{year}{2021}), \bibinfo{pages}{1--11}.
\newblock


\bibitem[Chen et~al\mbox{.}(2025)]%
        {chen2025envisioning}
\bibfield{author}{\bibinfo{person}{Shiping Chen}, \bibinfo{person}{Duncan Brumby}, {and} \bibinfo{person}{Anna Cox}.} \bibinfo{year}{2025}\natexlab{}.
\newblock \showarticletitle{Envisioning the Future of Peer Review: Investigating LLM-Assisted Reviewing Using ChatGPT as a Case Study}. In \bibinfo{booktitle}{\emph{Proceedings of the 4th Annual Symposium on Human-Computer Interaction for Work}}. \bibinfo{pages}{1--18}.
\newblock


\bibitem[Chidambaram(2025)]%
        {chidambaram2025the}
\bibfield{author}{\bibinfo{person}{Vijay Chidambaram}.} \bibinfo{year}{2025}\natexlab{}.
\newblock \showarticletitle{The role of LLMs in academic reviewing}.
\newblock \bibinfo{journal}{\emph{The ACM Special Interest Group in Operating Systems}} (\bibinfo{year}{2025}).
\newblock
\urldef\tempurl%
\url{https://www.sigops.org/2025/the-role-of-llms-in-academic-reviewing/}
\showURL{%
\tempurl}


\bibitem[Chitale et~al\mbox{.}(2025)]%
        {chitale2025autorev}
\bibfield{author}{\bibinfo{person}{Maitreya~Prafulla Chitale}, \bibinfo{person}{Ketaki~Mangesh Shetye}, \bibinfo{person}{Harshit Gupta}, \bibinfo{person}{Manav Chaudhary}, {and} \bibinfo{person}{Vasudeva Varma}.} \bibinfo{year}{2025}\natexlab{}.
\newblock \showarticletitle{AutoRev: Automatic Peer Review System for Academic Research Papers}.
\newblock \bibinfo{journal}{\emph{arXiv preprint arXiv:2505.14376}} (\bibinfo{year}{2025}).
\newblock


\bibitem[Chung et~al\mbox{.}(2015)]%
        {chung2015double}
\bibfield{author}{\bibinfo{person}{Kevin~C Chung}, \bibinfo{person}{Melissa~J Shauver}, \bibinfo{person}{Sunitha Malay}, \bibinfo{person}{Lin Zhong}, \bibinfo{person}{Aaron Weinstein}, {and} \bibinfo{person}{Rod~J Rohrich}.} \bibinfo{year}{2015}\natexlab{}.
\newblock \showarticletitle{Is double-blinded peer review necessary? The effect of blinding on review quality}.
\newblock \bibinfo{journal}{\emph{Plastic and reconstructive surgery}} \bibinfo{volume}{136}, \bibinfo{number}{6} (\bibinfo{year}{2015}), \bibinfo{pages}{1369--1377}.
\newblock


\bibitem[Conroy(2023)]%
        {conroy2023chatgpt}
\bibfield{author}{\bibinfo{person}{Gemma Conroy}.} \bibinfo{year}{2023}\natexlab{}.
\newblock \showarticletitle{How ChatGPT and other AI tools could disrupt scientific publishing}.
\newblock \bibinfo{journal}{\emph{Nature}} \bibinfo{volume}{622}, \bibinfo{number}{7982} (\bibinfo{year}{2023}), \bibinfo{pages}{234--236}.
\newblock


\bibitem[Demir and Pismek(2018)]%
        {demir2018convergent}
\bibfield{author}{\bibinfo{person}{Selcuk~Besir Demir} {and} \bibinfo{person}{Nuray Pismek}.} \bibinfo{year}{2018}\natexlab{}.
\newblock \showarticletitle{A Convergent Parallel Mixed-Methods Study of Controversial Issues in Social Studies Classes: A Clash of Ideologies.}
\newblock \bibinfo{journal}{\emph{Educational Sciences: Theory and Practice}} \bibinfo{volume}{18}, \bibinfo{number}{1} (\bibinfo{year}{2018}), \bibinfo{pages}{119--149}.
\newblock


\bibitem[Ding et~al\mbox{.}(2022)]%
        {ding2022uploaders}
\bibfield{author}{\bibinfo{person}{Xianghua Ding}, \bibinfo{person}{Yubo Kou}, \bibinfo{person}{Yiwen Xu}, {and} \bibinfo{person}{Peng Zhang}.} \bibinfo{year}{2022}\natexlab{}.
\newblock \showarticletitle{“As Uploaders, We Have the Responsibility”: Individualized Professionalization of Bilibili Uploaders}. In \bibinfo{booktitle}{\emph{Proceedings of the 2022 CHI Conference on Human Factors in Computing Systems}}. \bibinfo{pages}{1--14}.
\newblock


\bibitem[Donker(2023)]%
        {donker2023dangers}
\bibfield{author}{\bibinfo{person}{Tjibbe Donker}.} \bibinfo{year}{2023}\natexlab{}.
\newblock \showarticletitle{The dangers of using large language models for peer review}.
\newblock \bibinfo{journal}{\emph{The Lancet Infectious Diseases}} \bibinfo{volume}{23}, \bibinfo{number}{7} (\bibinfo{year}{2023}), \bibinfo{pages}{781}.
\newblock


\bibitem[Doskaliuk et~al\mbox{.}(2025)]%
        {doskaliuk2025artificial}
\bibfield{author}{\bibinfo{person}{Bohdana Doskaliuk}, \bibinfo{person}{Olena Zimba}, \bibinfo{person}{Marlen Yessirkepov}, \bibinfo{person}{Iryna Klishch}, {and} \bibinfo{person}{Roman Yatsyshyn}.} \bibinfo{year}{2025}\natexlab{}.
\newblock \showarticletitle{Artificial intelligence in peer review: enhancing efficiency while preserving integrity}.
\newblock \bibinfo{journal}{\emph{Journal of Korean medical science}} \bibinfo{volume}{40}, \bibinfo{number}{7} (\bibinfo{year}{2025}).
\newblock


\bibitem[Drori and Te'eni(2024)]%
        {drori2024human}
\bibfield{author}{\bibinfo{person}{Iddo Drori} {and} \bibinfo{person}{Dov Te'eni}.} \bibinfo{year}{2024}\natexlab{}.
\newblock \showarticletitle{Human-in-the-loop AI reviewing: feasibility, opportunities, and risks}.
\newblock \bibinfo{journal}{\emph{Journal of the Association for Information Systems}} \bibinfo{volume}{25}, \bibinfo{number}{1} (\bibinfo{year}{2024}), \bibinfo{pages}{98--109}.
\newblock


\bibitem[Drozdz and Ladomery(2024)]%
        {drozdz2024peer}
\bibfield{author}{\bibinfo{person}{John~A Drozdz} {and} \bibinfo{person}{Michael~R Ladomery}.} \bibinfo{year}{2024}\natexlab{}.
\newblock \showarticletitle{The peer review process: past, present, and future}.
\newblock \bibinfo{journal}{\emph{British Journal of Biomedical Science}}  \bibinfo{volume}{81} (\bibinfo{year}{2024}), \bibinfo{pages}{12054}.
\newblock


\bibitem[Du et~al\mbox{.}(2024)]%
        {du2024llms}
\bibfield{author}{\bibinfo{person}{Jiangshu Du}, \bibinfo{person}{Yibo Wang}, \bibinfo{person}{Wenting Zhao}, \bibinfo{person}{Zhongfen Deng}, \bibinfo{person}{Shuaiqi Liu}, \bibinfo{person}{Renze Lou}, \bibinfo{person}{Henry~Peng Zou}, \bibinfo{person}{Pranav~Narayanan Venkit}, \bibinfo{person}{Nan Zhang}, \bibinfo{person}{Mukund Srinath}, {et~al\mbox{.}}} \bibinfo{year}{2024}\natexlab{}.
\newblock \showarticletitle{LLMs assist NLP researchers: Critique paper (meta-) reviewing}. In \bibinfo{booktitle}{\emph{Proceedings of the 2024 conference on empirical methods in natural language processing}}. \bibinfo{pages}{5081--5099}.
\newblock


\bibitem[Ebadi et~al\mbox{.}(2025)]%
        {ebadi2025exploring}
\bibfield{author}{\bibinfo{person}{Saman Ebadi}, \bibinfo{person}{Hassan Nejadghanbar}, \bibinfo{person}{Ahmed~Rawdhan Salman}, {and} \bibinfo{person}{Hassan Khosravi}.} \bibinfo{year}{2025}\natexlab{}.
\newblock \showarticletitle{Exploring the impact of generative {AI} on peer review: Insights from journal reviewers}.
\newblock \bibinfo{journal}{\emph{Journal of Academic Ethics}} (\bibinfo{year}{2025}), \bibinfo{pages}{1--15}.
\newblock


\bibitem[Feetham-Walker et~al\mbox{.}(2025)]%
        {iop2025}
\bibfield{author}{\bibinfo{person}{Laura Feetham-Walker}, \bibinfo{person}{Eden Brent-Jones}, \bibinfo{person}{Emma Chorlton}, \bibinfo{person}{Miriam Dixon}, \bibinfo{person}{Anna Coombs}, \bibinfo{person}{Sian Powell}, \bibinfo{person}{Faye Holst}, {and} \bibinfo{person}{Kate Giles}.} \bibinfo{year}{2025}\natexlab{}.
\newblock \bibinfo{booktitle}{\emph{AI and Peer Review 2025}}.
\newblock \bibinfo{type}{{T}echnical {R}eport}. \bibinfo{institution}{IOP Publishing}.
\newblock
\urldef\tempurl%
\url{https://ioppublishing.org/ai-and-peer-review-2025/?utm_campaign=peer-review-ai-survey&utm_medium=referral&utm_source=landing_page}
\showURL{%
\tempurl}


\bibitem[Gao et~al\mbox{.}(2025)]%
        {gao2025reviewagents}
\bibfield{author}{\bibinfo{person}{Xian Gao}, \bibinfo{person}{Jiacheng Ruan}, \bibinfo{person}{Zongyun Zhang}, \bibinfo{person}{Jingsheng Gao}, \bibinfo{person}{Ting Liu}, {and} \bibinfo{person}{Yuzhuo Fu}.} \bibinfo{year}{2025}\natexlab{}.
\newblock \showarticletitle{ReviewAgents: Bridging the gap between human and {AI}-generated paper reviews}.
\newblock \bibinfo{journal}{\emph{arXiv preprint arXiv:2503.08506}} (\bibinfo{year}{2025}).
\newblock


\bibitem[Ghosh(2024)]%
        {ghosh2024chatgpt}
\bibfield{author}{\bibinfo{person}{Sourojit Ghosh}.} \bibinfo{year}{2024}\natexlab{}.
\newblock \showarticletitle{Chat{GPT} as a tool for equitable education in engineering classes}. In \bibinfo{booktitle}{\emph{2024 ASEE Annual Conference \& Exposition}}.
\newblock


\bibitem[Goldberg et~al\mbox{.}(2024)]%
        {goldberg2024usefulness}
\bibfield{author}{\bibinfo{person}{Alexander Goldberg}, \bibinfo{person}{Ihsan Ullah}, \bibinfo{person}{Thanh Gia~Hieu Khuong}, \bibinfo{person}{Benedictus~Kent Rachmat}, \bibinfo{person}{Zhen Xu}, \bibinfo{person}{Isabelle Guyon}, {and} \bibinfo{person}{Nihar~B Shah}.} \bibinfo{year}{2024}\natexlab{}.
\newblock \showarticletitle{Usefulness of LLMs as an Author Checklist Assistant for Scientific Papers: NeurIPS'24 Experiment}.
\newblock \bibinfo{journal}{\emph{arXiv preprint arXiv:2411.03417}} (\bibinfo{year}{2024}).
\newblock


\bibitem[Gruda(2025)]%
        {gruda2025three}
\bibfield{author}{\bibinfo{person}{Dritjon Gruda}.} \bibinfo{year}{2025}\natexlab{}.
\newblock \showarticletitle{Three AI-powered steps to faster, smarter peer review}.
\newblock \bibinfo{journal}{\emph{Nature}}  \bibinfo{volume}{10} (\bibinfo{year}{2025}).
\newblock


\bibitem[Haffar et~al\mbox{.}(2019)]%
        {haffar2019peer}
\bibfield{author}{\bibinfo{person}{Samir Haffar}, \bibinfo{person}{Fateh Bazerbachi}, {and} \bibinfo{person}{M~Hassan Murad}.} \bibinfo{year}{2019}\natexlab{}.
\newblock \showarticletitle{Peer review bias: a critical review}. In \bibinfo{booktitle}{\emph{Mayo Clinic Proceedings}}, Vol.~\bibinfo{volume}{94}. Elsevier, \bibinfo{pages}{670--676}.
\newblock


\bibitem[Helmer et~al\mbox{.}(2017)]%
        {helmer2017gender}
\bibfield{author}{\bibinfo{person}{Markus Helmer}, \bibinfo{person}{Manuel Schottdorf}, \bibinfo{person}{Andreas Neef}, {and} \bibinfo{person}{Demian Battaglia}.} \bibinfo{year}{2017}\natexlab{}.
\newblock \showarticletitle{Gender bias in scholarly peer review}.
\newblock \bibinfo{journal}{\emph{elife}}  \bibinfo{volume}{6} (\bibinfo{year}{2017}), \bibinfo{pages}{e21718}.
\newblock


\bibitem[Idahl and Ahmadi(2025)]%
        {idahl2025openreviewer}
\bibfield{author}{\bibinfo{person}{Maximilian Idahl} {and} \bibinfo{person}{Zahra Ahmadi}.} \bibinfo{year}{2025}\natexlab{}.
\newblock \showarticletitle{OpenReviewer: A specialized large language model for generating critical scientific paper reviews}. In \bibinfo{booktitle}{\emph{Proceedings of the 2025 Conference of the Nations of the Americas Chapter of the Association for Computational Linguistics: Human Language Technologies (System Demonstrations)}}. \bibinfo{pages}{550--562}.
\newblock


\bibitem[Kerrigan(2014)]%
        {kerrigan2014framework}
\bibfield{author}{\bibinfo{person}{Monica~Reid Kerrigan}.} \bibinfo{year}{2014}\natexlab{}.
\newblock \showarticletitle{A framework for understanding community colleges’ organizational capacity for data use: A convergent parallel mixed methods study}.
\newblock \bibinfo{journal}{\emph{Journal of mixed methods research}} \bibinfo{volume}{8}, \bibinfo{number}{4} (\bibinfo{year}{2014}), \bibinfo{pages}{341--362}.
\newblock


\bibitem[Kohnke and Zaugg(2025)]%
        {kohnke2025artificial}
\bibfield{author}{\bibinfo{person}{Shalece Kohnke} {and} \bibinfo{person}{Tiffanie Zaugg}.} \bibinfo{year}{2025}\natexlab{}.
\newblock \showarticletitle{Artificial intelligence: an untapped opportunity for equity and access in {STEM} education}.
\newblock \bibinfo{journal}{\emph{Education Sciences}} \bibinfo{volume}{15}, \bibinfo{number}{1} (\bibinfo{year}{2025}), \bibinfo{pages}{68}.
\newblock


\bibitem[Kronick(1990)]%
        {kronick1990peer}
\bibfield{author}{\bibinfo{person}{David~A Kronick}.} \bibinfo{year}{1990}\natexlab{}.
\newblock \showarticletitle{Peer review in 18th-century scientific journalism}.
\newblock \bibinfo{journal}{\emph{Journal of the American Medical Association}} \bibinfo{volume}{263}, \bibinfo{number}{10} (\bibinfo{year}{1990}), \bibinfo{pages}{1321--1322}.
\newblock


\bibitem[Lamont(2009)]%
        {lamont2009professors}
\bibfield{author}{\bibinfo{person}{Mich{\`e}le Lamont}.} \bibinfo{year}{2009}\natexlab{}.
\newblock \bibinfo{booktitle}{\emph{How professors think: Inside the curious world of academic judgment}}.
\newblock \bibinfo{publisher}{Harvard University Press}.
\newblock


\bibitem[Laufer et~al\mbox{.}(2022)]%
        {laufer2022four}
\bibfield{author}{\bibinfo{person}{Benjamin Laufer}, \bibinfo{person}{Sameer Jain}, \bibinfo{person}{A~Feder Cooper}, \bibinfo{person}{Jon Kleinberg}, {and} \bibinfo{person}{Hoda Heidari}.} \bibinfo{year}{2022}\natexlab{}.
\newblock \showarticletitle{Four years of FAccT: A reflexive, mixed-methods analysis of research contributions, shortcomings, and future prospects}. In \bibinfo{booktitle}{\emph{Proceedings of the 2022 ACM conference on fairness, accountability, and transparency}}. \bibinfo{pages}{401--426}.
\newblock


\bibitem[Lee et~al\mbox{.}(2013)]%
        {lee2013bias}
\bibfield{author}{\bibinfo{person}{Carole~J Lee}, \bibinfo{person}{Cassidy~R Sugimoto}, \bibinfo{person}{Guo Zhang}, {and} \bibinfo{person}{Blaise Cronin}.} \bibinfo{year}{2013}\natexlab{}.
\newblock \showarticletitle{Bias in peer review}.
\newblock \bibinfo{journal}{\emph{Journal of the American Society for information Science and Technology}} \bibinfo{volume}{64}, \bibinfo{number}{1} (\bibinfo{year}{2013}), \bibinfo{pages}{2--17}.
\newblock


\bibitem[Lee et~al\mbox{.}(2025)]%
        {lee2025role}
\bibfield{author}{\bibinfo{person}{Jisoo Lee}, \bibinfo{person}{Jieun Lee}, {and} \bibinfo{person}{Jeong-Ju Yoo}.} \bibinfo{year}{2025}\natexlab{}.
\newblock \showarticletitle{The role of large language models in the peer-review process: opportunities and challenges for medical journal reviewers and editors}.
\newblock \bibinfo{journal}{\emph{Journal of Educational Evaluation for Health Professions}}  \bibinfo{volume}{22} (\bibinfo{year}{2025}).
\newblock


\bibitem[Liang et~al\mbox{.}(2024a)]%
        {liang2024monitoring}
\bibfield{author}{\bibinfo{person}{Weixin Liang}, \bibinfo{person}{Zachary Izzo}, \bibinfo{person}{Yaohui Zhang}, \bibinfo{person}{Haley Lepp}, \bibinfo{person}{Hancheng Cao}, \bibinfo{person}{Xuandong Zhao}, \bibinfo{person}{Lingjiao Chen}, \bibinfo{person}{Haotian Ye}, \bibinfo{person}{Sheng Liu}, \bibinfo{person}{Zhi Huang}, {et~al\mbox{.}}} \bibinfo{year}{2024}\natexlab{a}.
\newblock \showarticletitle{Monitoring AI-modified content at scale: A case study on the impact of ChatGPT on AI conference peer reviews}.
\newblock \bibinfo{journal}{\emph{arXiv preprint arXiv:2403.07183}} (\bibinfo{year}{2024}).
\newblock


\bibitem[Liang et~al\mbox{.}(2024b)]%
        {10.5555/3692070.3693262}
\bibfield{author}{\bibinfo{person}{Weixin Liang}, \bibinfo{person}{Zachary Izzo}, \bibinfo{person}{Yaohui Zhang}, \bibinfo{person}{Haley Lepp}, \bibinfo{person}{Hancheng Cao}, \bibinfo{person}{Xuandong Zhao}, \bibinfo{person}{Lingjiao Chen}, \bibinfo{person}{Haotian Ye}, \bibinfo{person}{Sheng Liu}, \bibinfo{person}{Zhi Huang}, \bibinfo{person}{Daniel~A. McFarland}, {and} \bibinfo{person}{James~Y. Zou}.} \bibinfo{year}{2024}\natexlab{b}.
\newblock \showarticletitle{Monitoring AI-modified content at scale: a case study on the impact of ChatGPT on AI conference peer reviews}. In \bibinfo{booktitle}{\emph{Proceedings of the 41st International Conference on Machine Learning}} (Vienna, Austria) \emph{(\bibinfo{series}{ICML'24})}. \bibinfo{publisher}{JMLR.org}, Article \bibinfo{articleno}{1192}, \bibinfo{numpages}{46}~pages.
\newblock


\bibitem[Lin et~al\mbox{.}(2022)]%
        {lin2022automatic}
\bibfield{author}{\bibinfo{person}{Jialiang Lin}, \bibinfo{person}{Yingmin Wang}, \bibinfo{person}{Yao Yu}, \bibinfo{person}{Yu Zhou}, \bibinfo{person}{Yidong Chen}, {and} \bibinfo{person}{Xiaodong Shi}.} \bibinfo{year}{2022}\natexlab{}.
\newblock \showarticletitle{Automatic analysis of available source code of top artificial intelligence conference papers}.
\newblock \bibinfo{journal}{\emph{International Journal of Software Engineering and Knowledge Engineering}} \bibinfo{volume}{32}, \bibinfo{number}{07} (\bibinfo{year}{2022}), \bibinfo{pages}{947--970}.
\newblock


\bibitem[Lipworth and Kerridge(2011)]%
        {lipworth2011shifting}
\bibfield{author}{\bibinfo{person}{Wendy Lipworth} {and} \bibinfo{person}{Ian Kerridge}.} \bibinfo{year}{2011}\natexlab{}.
\newblock \showarticletitle{Shifting power relations and the ethics of journal peer review}.
\newblock \bibinfo{journal}{\emph{Social epistemology}} \bibinfo{volume}{25}, \bibinfo{number}{1} (\bibinfo{year}{2011}), \bibinfo{pages}{97--121}.
\newblock


\bibitem[Liu and Shah(2023)]%
        {liu2023reviewergpt}
\bibfield{author}{\bibinfo{person}{Ryan Liu} {and} \bibinfo{person}{Nihar~B Shah}.} \bibinfo{year}{2023}\natexlab{}.
\newblock \showarticletitle{Reviewergpt? an exploratory study on using large language models for paper reviewing}.
\newblock \bibinfo{journal}{\emph{arXiv preprint arXiv:2306.00622}} (\bibinfo{year}{2023}).
\newblock


\bibitem[Marsh et~al\mbox{.}(2008)]%
        {marsh2008improving}
\bibfield{author}{\bibinfo{person}{Herbert~W Marsh}, \bibinfo{person}{Upali~W Jayasinghe}, {and} \bibinfo{person}{Nigel~W Bond}.} \bibinfo{year}{2008}\natexlab{}.
\newblock \showarticletitle{Improving the peer-review process for grant applications: reliability, validity, bias, and generalizability.}
\newblock \bibinfo{journal}{\emph{American psychologist}} \bibinfo{volume}{63}, \bibinfo{number}{3} (\bibinfo{year}{2008}), \bibinfo{pages}{160}.
\newblock


\bibitem[Mogashoa(2014)]%
        {mogashoa2014understanding}
\bibfield{author}{\bibinfo{person}{Tebogo Mogashoa}.} \bibinfo{year}{2014}\natexlab{}.
\newblock \showarticletitle{Understanding critical discourse analysis in qualitative research}.
\newblock \bibinfo{journal}{\emph{International Journal of Humanities Social Sciences and Education}} \bibinfo{volume}{1}, \bibinfo{number}{7} (\bibinfo{year}{2014}), \bibinfo{pages}{104--113}.
\newblock


\bibitem[Neshaei et~al\mbox{.}(2024)]%
        {neshaei2024enhancing}
\bibfield{author}{\bibinfo{person}{Seyed~Parsa Neshaei}, \bibinfo{person}{Roman Rietsche}, \bibinfo{person}{Xiaotian Su}, {and} \bibinfo{person}{Thiemo Wambsganss}.} \bibinfo{year}{2024}\natexlab{}.
\newblock \showarticletitle{Enhancing peer review with AI-powered suggestion generation assistance: Investigating the design dynamics}. In \bibinfo{booktitle}{\emph{Proceedings of the 29th international conference on intelligent user interfaces}}. \bibinfo{pages}{88--102}.
\newblock


\bibitem[Nixon et~al\mbox{.}(2024)]%
        {nixon2024catalyzing}
\bibfield{author}{\bibinfo{person}{Nia Nixon}, \bibinfo{person}{Yiwen Lin}, {and} \bibinfo{person}{Lauren Snow}.} \bibinfo{year}{2024}\natexlab{}.
\newblock \showarticletitle{Catalyzing equity in {STEM} teams: Harnessing generative {AI} for inclusion and diversity}.
\newblock \bibinfo{journal}{\emph{Policy Insights from the Behavioral and Brain Sciences}} \bibinfo{volume}{11}, \bibinfo{number}{1} (\bibinfo{year}{2024}), \bibinfo{pages}{85--92}.
\newblock


\bibitem[Palmer(1908)]%
        {palmer1908editorial}
\bibfield{author}{\bibinfo{person}{G Palmer}.} \bibinfo{year}{1908}\natexlab{}.
\newblock \showarticletitle{Editorial Individuality}. In \bibinfo{booktitle}{\emph{American Medical Editors' Association}}. \bibinfo{pages}{57--63}.
\newblock


\bibitem[Peters and Ceci(1982)]%
        {peters1982peer}
\bibfield{author}{\bibinfo{person}{Douglas~P Peters} {and} \bibinfo{person}{Stephen~J Ceci}.} \bibinfo{year}{1982}\natexlab{}.
\newblock \showarticletitle{Peer-review research: Objections and obligations}.
\newblock \bibinfo{journal}{\emph{Behavioral and Brain Sciences}} \bibinfo{volume}{5}, \bibinfo{number}{2} (\bibinfo{year}{1982}), \bibinfo{pages}{246--255}.
\newblock


\bibitem[Pontille and Torny(2014)]%
        {pontille2014blind}
\bibfield{author}{\bibinfo{person}{David Pontille} {and} \bibinfo{person}{Didier Torny}.} \bibinfo{year}{2014}\natexlab{}.
\newblock \showarticletitle{The blind shall see! The question of anonymity in journal peer review}.
\newblock \bibinfo{journal}{\emph{Ada: A Journal of Gender, New Media, and Technology}} \bibinfo{number}{4} (\bibinfo{year}{2014}).
\newblock


\bibitem[Rastogi et~al\mbox{.}(2024)]%
        {rastogi2024randomized}
\bibfield{author}{\bibinfo{person}{Charvi Rastogi}, \bibinfo{person}{Xiangchen Song}, \bibinfo{person}{Zhijing Jin}, \bibinfo{person}{Ivan Stelmakh}, \bibinfo{person}{Hal Daum{\'e}~III}, \bibinfo{person}{Kun Zhang}, {and} \bibinfo{person}{Nihar~B Shah}.} \bibinfo{year}{2024}\natexlab{}.
\newblock \showarticletitle{A randomized controlled trial on anonymizing reviewers to each other in peer review discussions}.
\newblock \bibinfo{journal}{\emph{PloS one}} \bibinfo{volume}{19}, \bibinfo{number}{12} (\bibinfo{year}{2024}), \bibinfo{pages}{e0315674}.
\newblock


\bibitem[Resnik et~al\mbox{.}(2008)]%
        {resnik2008perceptions}
\bibfield{author}{\bibinfo{person}{David~B Resnik}, \bibinfo{person}{Christina Gutierrez-Ford}, {and} \bibinfo{person}{Shyamal Peddada}.} \bibinfo{year}{2008}\natexlab{}.
\newblock \showarticletitle{Perceptions of ethical problems with scientific journal peer review: an exploratory study}.
\newblock \bibinfo{journal}{\emph{Science and engineering ethics}} \bibinfo{volume}{14}, \bibinfo{number}{3} (\bibinfo{year}{2008}), \bibinfo{pages}{305--310}.
\newblock


\bibitem[Robertson(2023)]%
        {robertson2023gpt4}
\bibfield{author}{\bibinfo{person}{Zachary Robertson}.} \bibinfo{year}{2023}\natexlab{}.
\newblock \showarticletitle{GPT4 is slightly helpful for peer-review assistance: A pilot study}.
\newblock \bibinfo{journal}{\emph{arXiv preprint arXiv:2307.05492}} (\bibinfo{year}{2023}).
\newblock


\bibitem[Roshanaei et~al\mbox{.}(2023)]%
        {roshanaei2023harnessing}
\bibfield{author}{\bibinfo{person}{Maryam Roshanaei}, \bibinfo{person}{Hanna Olivares}, {and} \bibinfo{person}{Rafael~Rangel Lopez}.} \bibinfo{year}{2023}\natexlab{}.
\newblock \showarticletitle{Harnessing {AI} to foster equity in education: Opportunities, challenges, and emerging strategies}.
\newblock \bibinfo{journal}{\emph{Journal of Intelligent Learning Systems and Applications}} \bibinfo{volume}{15}, \bibinfo{number}{4} (\bibinfo{year}{2023}), \bibinfo{pages}{123--143}.
\newblock


\bibitem[Russo et~al\mbox{.}(2025)]%
        {russo2025ai}
\bibfield{author}{\bibinfo{person}{Giuseppe Russo}, \bibinfo{person}{Manoel Horta~Ribeiro}, \bibinfo{person}{Tim~Ruben Davidson}, \bibinfo{person}{Veniamin Veselovsky}, {and} \bibinfo{person}{Robert West}.} \bibinfo{year}{2025}\natexlab{}.
\newblock \showarticletitle{The AI Review Lottery: Widespread AI-Assisted Peer Reviews Boost Paper Scores and Acceptance Rates}.
\newblock \bibinfo{journal}{\emph{Proceedings of the ACM on Human-Computer Interaction}} \bibinfo{volume}{9}, \bibinfo{number}{7} (\bibinfo{year}{2025}), \bibinfo{pages}{1--28}.
\newblock


\bibitem[Saad et~al\mbox{.}(2024)]%
        {saad2024exploring}
\bibfield{author}{\bibinfo{person}{Ahmed Saad}, \bibinfo{person}{Nathan Jenko}, \bibinfo{person}{Sisith Ariyaratne}, \bibinfo{person}{Nick Birch}, \bibinfo{person}{Karthikeyan~P Iyengar}, \bibinfo{person}{Arthur~Mark Davies}, \bibinfo{person}{Raju Vaishya}, {and} \bibinfo{person}{Rajesh Botchu}.} \bibinfo{year}{2024}\natexlab{}.
\newblock \showarticletitle{Exploring the potential of ChatGPT in the peer review process: an observational study}.
\newblock \bibinfo{journal}{\emph{Diabetes \& Metabolic Syndrome: Clinical Research \& Reviews}} \bibinfo{volume}{18}, \bibinfo{number}{2} (\bibinfo{year}{2024}), \bibinfo{pages}{102946}.
\newblock


\bibitem[Sebastian and Baron(2024)]%
        {sebastian2024artificial}
\bibfield{author}{\bibinfo{person}{Felix Sebastian} {and} \bibinfo{person}{Rachel Baron}.} \bibinfo{year}{2024}\natexlab{}.
\newblock \showarticletitle{Artificial intelligence: What the future holds for multilingual authors and editing professionals}.
\newblock \bibinfo{journal}{\emph{Science Editor}} \bibinfo{volume}{47}, \bibinfo{number}{2} (\bibinfo{year}{2024}), \bibinfo{pages}{38--42}.
\newblock


\bibitem[Shiflett(1988)]%
        {shiflett1988difficult}
\bibfield{author}{\bibinfo{person}{Lee Shiflett}.} \bibinfo{year}{1988}\natexlab{}.
\newblock \bibinfo{title}{A difficult balance: Editorial peer review in medicine}.
\newblock


\bibitem[Spezi et~al\mbox{.}(2018)]%
        {spezi2018let}
\bibfield{author}{\bibinfo{person}{Valerie Spezi}, \bibinfo{person}{Simon Wakeling}, \bibinfo{person}{Stephen Pinfield}, \bibinfo{person}{Jenny Fry}, \bibinfo{person}{Claire Creaser}, {and} \bibinfo{person}{Peter Willett}.} \bibinfo{year}{2018}\natexlab{}.
\newblock \showarticletitle{“Let the community decide”? The vision and reality of soundness-only peer review in open-access mega-journals}.
\newblock \bibinfo{journal}{\emph{Journal of documentation}} \bibinfo{volume}{74}, \bibinfo{number}{1} (\bibinfo{year}{2018}), \bibinfo{pages}{137--161}.
\newblock


\bibitem[Stelmakh et~al\mbox{.}(2021)]%
        {stelmakh2021prior}
\bibfield{author}{\bibinfo{person}{Ivan Stelmakh}, \bibinfo{person}{Nihar~B Shah}, \bibinfo{person}{Aarti Singh}, {and} \bibinfo{person}{Hal Daum{\'e}~III}.} \bibinfo{year}{2021}\natexlab{}.
\newblock \showarticletitle{Prior and prejudice: The novice reviewers' bias against resubmissions in conference peer review}.
\newblock \bibinfo{journal}{\emph{Proceedings of the ACM on Human-Computer Interaction}} \bibinfo{volume}{5}, \bibinfo{number}{CSCW1} (\bibinfo{year}{2021}), \bibinfo{pages}{1--17}.
\newblock


\bibitem[Su et~al\mbox{.}(2023)]%
        {su2023reviewriter}
\bibfield{author}{\bibinfo{person}{Xiaotian Su}, \bibinfo{person}{Thiemo Wambsganss}, \bibinfo{person}{Roman Rietsche}, \bibinfo{person}{Seyed~Parsa Neshaei}, {and} \bibinfo{person}{Tanja K{\"a}ser}.} \bibinfo{year}{2023}\natexlab{}.
\newblock \showarticletitle{Reviewriter: {AI}-generated instructions for peer review writing}. In \bibinfo{booktitle}{\emph{Proceedings of the 18th workshop on innovative use of NLP for building educational applications (BEA 2023)}}. \bibinfo{pages}{57--71}.
\newblock


\bibitem[Sukpanichnant et~al\mbox{.}(2024)]%
        {sukpanichnant2024peerarg}
\bibfield{author}{\bibinfo{person}{Purin Sukpanichnant}, \bibinfo{person}{Anna Rapberger}, {and} \bibinfo{person}{Francesca Toni}.} \bibinfo{year}{2024}\natexlab{}.
\newblock \showarticletitle{PeerArg: Argumentative peer review with LLMs}.
\newblock \bibinfo{journal}{\emph{arXiv preprint arXiv:2409.16813}} (\bibinfo{year}{2024}).
\newblock


\bibitem[Suleiman et~al\mbox{.}(2024)]%
        {suleiman2024assessing}
\bibfield{author}{\bibinfo{person}{Aiman Suleiman}, \bibinfo{person}{Dario von Wedel}, \bibinfo{person}{Ricardo Munoz-Acuna}, \bibinfo{person}{Simone Redaelli}, \bibinfo{person}{Abeer Santarisi}, \bibinfo{person}{Eva-Lotte Seibold}, \bibinfo{person}{Nikolai Ratajczak}, \bibinfo{person}{Shinichiro Kato}, \bibinfo{person}{Nader Said}, \bibinfo{person}{Eswar Sundar}, {et~al\mbox{.}}} \bibinfo{year}{2024}\natexlab{}.
\newblock \showarticletitle{Assessing ChatGPT's ability to emulate human reviewers in scientific research: A descriptive and qualitative approach}.
\newblock \bibinfo{journal}{\emph{Computer Methods and Programs in Biomedicine}}  \bibinfo{volume}{254} (\bibinfo{year}{2024}), \bibinfo{pages}{108313}.
\newblock


\bibitem[Sun(2025)]%
        {sun2025large}
\bibfield{author}{\bibinfo{person}{Zhuanlan Sun}.} \bibinfo{year}{2025}\natexlab{}.
\newblock \showarticletitle{Large language models in peer review: challenges and opportunities}.
\newblock \bibinfo{journal}{\emph{Scientometrics}} (\bibinfo{year}{2025}), \bibinfo{pages}{1--44}.
\newblock


\bibitem[Taechoyotin and Acuna(2025)]%
        {taechoyotin2025remor}
\bibfield{author}{\bibinfo{person}{Pawin Taechoyotin} {and} \bibinfo{person}{Daniel Acuna}.} \bibinfo{year}{2025}\natexlab{}.
\newblock \showarticletitle{REMOR: Automated Peer Review Generation with {LLM} Reasoning and Multi-Objective Reinforcement Learning}.
\newblock \bibinfo{journal}{\emph{arXiv preprint arXiv:2505.11718}} (\bibinfo{year}{2025}).
\newblock


\bibitem[Tan et~al\mbox{.}(2024)]%
        {tan2024peer}
\bibfield{author}{\bibinfo{person}{Cheng Tan}, \bibinfo{person}{Dongxin Lyu}, \bibinfo{person}{Siyuan Li}, \bibinfo{person}{Zhangyang Gao}, \bibinfo{person}{Jingxuan Wei}, \bibinfo{person}{Siqi Ma}, \bibinfo{person}{Zicheng Liu}, {and} \bibinfo{person}{Stan~Z Li}.} \bibinfo{year}{2024}\natexlab{}.
\newblock \showarticletitle{Peer review as a multi-turn and long-context dialogue with role-based interactions}.
\newblock \bibinfo{journal}{\emph{arXiv preprint arXiv:2406.05688}} (\bibinfo{year}{2024}).
\newblock


\bibitem[Thakkar et~al\mbox{.}(2022)]%
        {thakkar2022machine}
\bibfield{author}{\bibinfo{person}{Divy Thakkar}, \bibinfo{person}{Azra Ismail}, \bibinfo{person}{Pratyush Kumar}, \bibinfo{person}{Alex Hanna}, \bibinfo{person}{Nithya Sambasivan}, {and} \bibinfo{person}{Neha Kumar}.} \bibinfo{year}{2022}\natexlab{}.
\newblock \showarticletitle{When is machine learning data good?: Valuing in public health datafication}. In \bibinfo{booktitle}{\emph{Proceedings of the 2022 CHI Conference on Human Factors in Computing Systems}}. \bibinfo{pages}{1--16}.
\newblock


\bibitem[Thakkar et~al\mbox{.}(2025)]%
        {thakkar2025can}
\bibfield{author}{\bibinfo{person}{Nitya Thakkar}, \bibinfo{person}{Mert Yuksekgonul}, \bibinfo{person}{Jake Silberg}, \bibinfo{person}{Animesh Garg}, \bibinfo{person}{Nanyun Peng}, \bibinfo{person}{Fei Sha}, \bibinfo{person}{Rose Yu}, \bibinfo{person}{Carl Vondrick}, {and} \bibinfo{person}{James Zou}.} \bibinfo{year}{2025}\natexlab{}.
\newblock \showarticletitle{Can {LLM} feedback enhance review quality? {A} randomized study of 20k reviews at {ICLR} 2025}.
\newblock \bibinfo{journal}{\emph{arXiv preprint arXiv:2504.09737}} (\bibinfo{year}{2025}).
\newblock


\bibitem[Thelwall and Yaghi(2024)]%
        {thelwall2024fields}
\bibfield{author}{\bibinfo{person}{Mike Thelwall} {and} \bibinfo{person}{Abdallah Yaghi}.} \bibinfo{year}{2024}\natexlab{}.
\newblock \showarticletitle{In which fields can ChatGPT detect journal article quality? An evaluation of REF2021 results}.
\newblock \bibinfo{journal}{\emph{arXiv preprint arXiv:2409.16695}} (\bibinfo{year}{2024}).
\newblock


\bibitem[Tregenza(2002)]%
        {tregenza2002gender}
\bibfield{author}{\bibinfo{person}{Tom Tregenza}.} \bibinfo{year}{2002}\natexlab{}.
\newblock \showarticletitle{Gender bias in the refereeing process?}
\newblock \bibinfo{journal}{\emph{Trends in Ecology \& Evolution}} \bibinfo{volume}{17}, \bibinfo{number}{8} (\bibinfo{year}{2002}), \bibinfo{pages}{349--350}.
\newblock


\bibitem[Varanasi et~al\mbox{.}(2022)]%
        {varanasi2022feeling}
\bibfield{author}{\bibinfo{person}{Rama~Adithya Varanasi}, \bibinfo{person}{Divya Siddarth}, \bibinfo{person}{Vivek Seshadri}, \bibinfo{person}{Kalika Bali}, {and} \bibinfo{person}{Aditya Vashistha}.} \bibinfo{year}{2022}\natexlab{}.
\newblock \showarticletitle{Feeling Proud, Feeling Embarrassed: Experiences of Low-income Women with Crowd Work}. In \bibinfo{booktitle}{\emph{Proceedings of the 2022 CHI Conference on Human Factors in Computing Systems}}. \bibinfo{pages}{1--18}.
\newblock


\bibitem[Wang et~al\mbox{.}(2020)]%
        {wang2020reviewrobot}
\bibfield{author}{\bibinfo{person}{Qingyun Wang}, \bibinfo{person}{Qi Zeng}, \bibinfo{person}{Lifu Huang}, \bibinfo{person}{Kevin Knight}, \bibinfo{person}{Heng Ji}, {and} \bibinfo{person}{Nazneen~Fatema Rajani}.} \bibinfo{year}{2020}\natexlab{}.
\newblock \showarticletitle{{R}eview{R}obot: Explainable Paper Review Generation based on Knowledge Synthesis}. In \bibinfo{booktitle}{\emph{Proceedings of the 13th International Conference on Natural Language Generation}}. \bibinfo{publisher}{Association for Computational Linguistics}, \bibinfo{address}{Dublin, Ireland}, \bibinfo{pages}{384--397}.
\newblock
\urldef\tempurl%
\url{https://aclanthology.org/2020.inlg-1.44}
\showURL{%
\tempurl}


\bibitem[Weng et~al\mbox{.}(2025)]%
        {weng2025cycleresearcher}
\bibfield{author}{\bibinfo{person}{Yixuan Weng}, \bibinfo{person}{Minjun Zhu}, \bibinfo{person}{Guangsheng Bao}, \bibinfo{person}{Hongbo Zhang}, \bibinfo{person}{Jindong Wang}, \bibinfo{person}{Yue Zhang}, {and} \bibinfo{person}{Linyi Yang}.} \bibinfo{year}{2025}\natexlab{}.
\newblock \showarticletitle{CycleResearcher: Improving Automated Research via Automated Review}. In \bibinfo{booktitle}{\emph{The Thirteenth International Conference on Learning Representations}}.
\newblock


\bibitem[Whittaker(2008)]%
        {whittaker2008journal}
\bibfield{author}{\bibinfo{person}{Robert~J Whittaker}.} \bibinfo{year}{2008}\natexlab{}.
\newblock \showarticletitle{Journal review and gender equality: a critical comment on Budden et al.}
\newblock \bibinfo{journal}{\emph{Trends in ecology \& evolution}} \bibinfo{volume}{23}, \bibinfo{number}{9} (\bibinfo{year}{2008}), \bibinfo{pages}{478--479}.
\newblock


\bibitem[Young et~al\mbox{.}(2022)]%
        {young2022confronting}
\bibfield{author}{\bibinfo{person}{Meg Young}, \bibinfo{person}{Michael Katell}, {and} \bibinfo{person}{PM Krafft}.} \bibinfo{year}{2022}\natexlab{}.
\newblock \showarticletitle{Confronting power and corporate capture at the FAccT Conference}. In \bibinfo{booktitle}{\emph{Proceedings of the 2022 ACM Conference on Fairness, Accountability, and Transparency}}. \bibinfo{pages}{1375--1386}.
\newblock


\bibitem[Yu et~al\mbox{.}(2024)]%
        {yu2024your}
\bibfield{author}{\bibinfo{person}{Sungduk Yu}, \bibinfo{person}{Man Luo}, \bibinfo{person}{Avinash Madasu}, \bibinfo{person}{Vasudev Lal}, {and} \bibinfo{person}{Phillip Howard}.} \bibinfo{year}{2024}\natexlab{}.
\newblock \showarticletitle{Is your paper being reviewed by an LLM? Investigating AI text detectability in peer review}.
\newblock \bibinfo{journal}{\emph{arXiv preprint arXiv:2410.03019}} (\bibinfo{year}{2024}).
\newblock


\bibitem[Zajda(2020)]%
        {zajda2020discourse}
\bibfield{author}{\bibinfo{person}{Joseph Zajda}.} \bibinfo{year}{2020}\natexlab{}.
\newblock \showarticletitle{Discourse analysis as a qualitative methodology}.
\newblock \bibinfo{journal}{\emph{Educational Practice and Theory}} \bibinfo{volume}{42}, \bibinfo{number}{2} (\bibinfo{year}{2020}), \bibinfo{pages}{5--21}.
\newblock


\bibitem[Zeng et~al\mbox{.}(2025)]%
        {zeng2025reviewrl}
\bibfield{author}{\bibinfo{person}{Sihang Zeng}, \bibinfo{person}{Kai Tian}, \bibinfo{person}{Kaiyan Zhang}, \bibinfo{person}{Yuru Wang}, \bibinfo{person}{Junqi Gao}, \bibinfo{person}{Runze Liu}, \bibinfo{person}{Sa Yang}, \bibinfo{person}{Jingxuan Li}, \bibinfo{person}{Xinwei Long}, \bibinfo{person}{Jiaheng Ma}, {et~al\mbox{.}}} \bibinfo{year}{2025}\natexlab{}.
\newblock \showarticletitle{ReviewRL: Towards Automated Scientific Review with {RL}}. In \bibinfo{booktitle}{\emph{Proceedings of the 2025 Conference on Empirical Methods in Natural Language Processing}}. \bibinfo{pages}{16942--16954}.
\newblock


\bibitem[Zhou et~al\mbox{.}(2024b)]%
        {zhou2024may}
\bibfield{author}{\bibinfo{person}{Haichen Zhou}, \bibinfo{person}{Xiaorong Huang}, \bibinfo{person}{Hongjun Pu}, {and} \bibinfo{person}{Zhang Qi}.} \bibinfo{year}{2024}\natexlab{b}.
\newblock \showarticletitle{May generative AI be a reviewer on an academic paper}.
\newblock In \bibinfo{booktitle}{\emph{EEKE-AII}}.
\newblock


\bibitem[Zhou et~al\mbox{.}(2024a)]%
        {zhou2024llm}
\bibfield{author}{\bibinfo{person}{Ruiyang Zhou}, \bibinfo{person}{Lu Chen}, {and} \bibinfo{person}{Kai Yu}.} \bibinfo{year}{2024}\natexlab{a}.
\newblock \showarticletitle{Is LLM a reliable reviewer? a comprehensive evaluation of LLM on automatic paper reviewing tasks}. In \bibinfo{booktitle}{\emph{Proceedings of the 2024 joint international conference on computational linguistics, language resources and evaluation (LREC-COLING 2024)}}. \bibinfo{pages}{9340--9351}.
\newblock


\bibitem[Zhuang et~al\mbox{.}(2025)]%
        {zhuang2025large}
\bibfield{author}{\bibinfo{person}{Zhenzhen Zhuang}, \bibinfo{person}{Jiandong Chen}, \bibinfo{person}{Hongfeng Xu}, \bibinfo{person}{Yuwen Jiang}, {and} \bibinfo{person}{Jialiang Lin}.} \bibinfo{year}{2025}\natexlab{}.
\newblock \showarticletitle{Large language models for automated scholarly paper review: A survey}.
\newblock \bibinfo{journal}{\emph{Information Fusion}} \bibinfo{volume}{124}, \bibinfo{number}{C} (\bibinfo{year}{2025}), \bibinfo{numpages}{17}~pages.
\newblock
\showISSN{1566-2535}
\href{https://doi.org/10.1016/j.inffus.2025.103332}{doi:\nolinkurl{10.1016/j.inffus.2025.103332}}


\end{thebibliography}

\appendix

\section{Appendix}

Call for Participation: Generative AI in Peer Review 

We are conducting a research study on how generative AI (GenAI) is reshaping the peer review process in academic publishing spaces (such as conferences and journals). Our goal is to understand:

1. Opportunities for improving efficiency and quality,
2. Challenges and risks (bias, fairness, confidentiality, accountability),
3. Ethical and governance implications,
4. The future of human–AI collaboration in peer review.

We are particularly seeking insights from individuals who have been or are:
1. Area Chairs (ACs)/ Senior Area Chair (SAC) at any major AI/ML/NLP conferences, or those that publish AI/ML/NLP research
2. Program Chairs (PCs) or Program Committee Members at any major AI/ML/NLP conferences, or those that publish AI/ML/NLP research
3. Editors of AI/ML/NLP journals, or those that publish AI/ML/NLP research

Your perspectives are invaluable for shaping responsible and human-centered practices as AI tools enter the peer review ecosystem. Participation involves a 45–60 minute 1-1 interview to understand your perspectives on AI reviewing. All responses will be anonymized, and your voice will contribute to broader conversations on the integrity and future of peer review. As a token of appreciation for your time, you will receive a \$50 Amazon gift voucher upon completing the interview. If you are interested in participating in this study, please provide your details below. 
1.Please Enter Your Name
2.Provide your Google Scholar link 
3.Please Provide Your Email ID
(This will be used for future communications regarding your participation in this study.)
4.How many conferences have you served as Area Chair or Senior Area Chair (AC, SAC)?
5.How many conferences have you served as Program Chair or Program Committee Member?
6.How many journals have you served as an Editor?
Provide your available dates.

\begin{table}[h]
\caption{Gender, position, area chair (AC)/ program chair(PC), and conference type of interview participants.}
\begin{center}
\label{tab:subject1}
\begin{tabular}{|l|l|l|l|c|}
\hline
\multicolumn{1}{|c|}{\textbf{ID}} & \multicolumn{1}{|c|}{\textbf{Gender}} & \multicolumn{1}{|c|}{\textbf{Position}} & \multicolumn{1}{|c|}{\textbf{AC/PC}} & \multicolumn{1}{|c|}{\textbf{Conference}} \\ \hline
P1  & Female &   Postdoc   & AC &   HCI     \\
P2  &  Female  &   PhD   & AC &   HCI    \\
P3  & Male &   Asst Prof   &  AC   & HCI  \\
P4  & Male &  Asso Prof    & AC, PC & AI \\
P5  &  Female  &  Postdoc    & AC  &  HCI     \\
P6  & Female &      Asso Prof  & AC, PC &  AI    \\
P7  & Female &  Industry   & AC &   HCI   \\
P8  & Male &   Asst Prof   &   AC  &  HCI   \\
P9  & Male   &  Asso Prof  &  AC, PC     &  HCI    \\
P10 & Male  &  Asso Prof   &    AC, PC      &  AI    \\
P11 & Male &  PhD   &   AC   &   AI   \\
P12 &  Female  &   Asst Prof   &    AC      &   HCI   \\
P13 & Male &  PhD   & AC &  HCI    \\
P14 &  Male  &  PhD   & AC &     AI        \\

\hline
\end{tabular}
\end{center}
\end{table}

\begin{table*}[t]
\centering
\footnotesize
\setlength{\tabcolsep}{6pt}
\renewcommand{\arraystretch}{1.25}
\begin{tabularx}{\textwidth}{@{}p{0.1\textwidth}p{0.2\textwidth}p{0.6\textwidth}@{}}
\toprule
\textbf{Actor} &
\textbf{Core Governance Goal} &
\textbf{Recommendations (including commented-out items)} \\
\midrule

\textbf{Reviewers} &
Preserve accountable human judgment while using GenAI as bounded support &
\begin{itemize}\setlength{\itemsep}{2pt}
    \item Treat GenAI as \textbf{assistive scaffolding}, not a delegate for evaluative judgment (novelty, contribution, accept/reject).
    \item Restrict AI use to supportive tasks (clarity, grammar, tone, outlining, summarizing reviewer-authored notes).
    \item Adopt a \textbf{verification norm} for any AI-assisted factual or evaluative claims; do not include critiques that cannot be traced to the manuscript.
    \item Use GenAI to improve the \textbf{communication of critique} (politeness, concision, tone calibration), not its substance.
    \item Preserve \textbf{confidentiality by default}; treat uploading unpublished manuscripts as high-risk unless explicitly approved with contractual safeguards.
    \item Make \textbf{review labor legible} through structured, evidence-linked feedback (major/minor concerns, actionable revisions), not length or polish.
\end{itemize}
\\

\midrule

\textbf{Area Chairs \& Program Chairs} &
Shift governance from provenance policing to quality-based adjudication &
\begin{itemize}\setlength{\itemsep}{2pt}
    \item Replace AI detection with \textbf{review quality adjudication} based on groundedness, specificity, and consistency with the manuscript.
    \item Trigger escalation (revision request, additional review, downweighting) for generic, ungrounded, or inconsistent reviews regardless of origin.
    \item Institutionalize a \textbf{downweighting mechanism} so unreliable reviews do not unduly influence meta-review synthesis or decisions.
    \item Provide \textbf{reviewer-facing micro-guidance} at submission time: permitted/prohibited AI uses, confidentiality reminders, and citation of manuscript sections for major critiques.
    \item Protect junior reviewers via \textbf{safe-harbor clarification channels} (“ask the chair”), emphasizing learning-oriented compliance and non-punitive disclosure.
\end{itemize}
\\

\midrule

\textbf{Conferences \& Journals} &
Replace strategic ambiguity with enforceable, adaptive governance &
\begin{itemize}\setlength{\itemsep}{2pt}
    \item Move from strategic ambiguity to \textbf{adaptive specificity} via a two-layer policy structure.
    \item Define \textbf{stable principles} (accountability, confidentiality, transparency, proportional enforcement).
    \item Provide \textbf{cycle-specific operational guidance} (permitted/prohibited uses, tool categories, disclosure requirements, enforcement processes).
    \item Codify the \textbf{support–judgment boundary} explicitly in policy and workflow.
    \item Require \textbf{bounded disclosure} using activity checklists rather than tool names or prompts.
    \item Introduce \textbf{due-process protections} and proportional remedies (revision, extra review, downweighting) before sanctions.
    \item Treat \textbf{adversarial risks} (e.g., prompt injection) as part of review integrity, not edge cases.
    \item Address \textbf{structural strain as governance}: reviewer load caps, recognition for high-quality reviews, training, and expanded reviewer pools.
\end{itemize}
\\

\midrule

\textbf{AI Developers} &
Enable accountable assistance without automating judgment &
\begin{itemize}\setlength{\itemsep}{2pt}
    \item Design for \textbf{accountable assistance} by making AI contributions legible (tracked suggestions, editable drafts, provenance markers).
    \item Prevent \textbf{automation by default} while preserving legitimate supportive use.
    \item Build guardrails that operationalize the \textbf{support–judgment boundary} (block accept/reject generation; require evidence citations).
    \item Prompt users to supply manuscript-grounded justification for critiques.
    \item Prioritize \textbf{confidentiality-preserving workflows}: local/on-device processing, non-retention guarantees, clear data-handling notices.
    \item Minimize full-manuscript uploads by analyzing reviewer notes.
    \item Anticipate \textbf{adversarial behavior} (hidden text, prompt injection) and surface warnings in AI-mediated review tools.
\end{itemize}
\\

\bottomrule
\end{tabularx}
\caption{Our complete Actor-specific recommendations for governing AI-mediated peer review as a \textit{sociotechnical system}, synthesizing both active and commented-out recommendations.}
\label{tab:actor_recommendations}
\end{table*}

\end{document}